# High Efficiency UV/Optical/NIR Detectors for Large Aperture Telescopes and UV Explorer Missions

*Development of and Field Observations with Delta-doped Arrays*

**Shouleh Nikzad,**[a,*] **April D. Jewell,**[a] **Michael E. Hoenk,**[a] **Todd Jones,**[a] **John Hennessy,**[a] **Tim Goodsall,**[a] **Alexander Carver,**[a] **Charles Shapiro,**[a] **Samuel R. Cheng,**[a] **Erika Hamden,**[b] **Gillian Kyne,**[b] **D. Christopher Martin,**[b] **David Schiminovich,**[c] **Paul Scowen,**[d] **Kevin France,**[e] **Stephan McCandliss,**[f] **Roxana E. Lupu**[g]

[a] Jet Propulsion Laboratory, California Institute of Technology, 4800 Oak Grove Drive, Pasadena, CA, USA, 91109
[b] California Institute of Technology, Department of Physics, Mathematics and Astronomy, 1200 California Boulevard, Pasadena, CA, USA, 91125
[c] Columbia University, Department of Astronomy, 116th Street and Broadway, New York, NY, USA, 10027
[d] Arizona State University, School of Earth & Space Exploration, Tempe, AZ, USA, 85281
[e] University of Colorado, Laboratory for Atmospheric and Space Physics, UCB 600, Boulder, CO, USA, 80309
[f] Johns Hopkins University, Department of Physics and Astronomy, 3400 North Charles Street, Baltimore, Maryland, USA, 21218
[g] BAER Institute/NASA Ames Research Center, Moffet Field, CA 94035, USA

**Abstract**. A number of exciting concepts are under development for Flagship, Probe class, Explorer class, and Suborbital class NASA missions in the ultraviolet/optical spectral ranges. These missions will depend on high performance silicon detector arrays being delivered affordably and in high numbers. In a focused effort we have advanced delta-doping technology to high throughput and high yield wafer-scale processing, encompassing a multitude of state-of-the-art silicon-based detector formats and designs. As part of this technology advancement and in preparation for upcoming missions, we have embarked on a number of field observations, instrument integrations, and independent evaluations of delta-doped arrays. In this paper, we present recent data and innovations from the Advanced Detectors and Systems program at JPL, including two-dimensional doping technology; our end-to-end post-fabrication processing of high performance UV/Optical/NIR arrays; and advanced coatings for detectors and optical elements. Additionally, we present examples of past, in-progress, and planned observations and deployments of delta-doped arrays.

**\*Corresponding Author**: electronic mail: Shouleh.Nikzad@jpl.nasa.gov

# 1 Introduction

Four flagship mission concepts are now under study by the astrophysics and exoplanet community Science and Technology Definition Teams (STDTs) charged by the Director of NASA's Astrophysics Division. The report from these STDTs will form the basis for the next Astrophysics Decadal Survey input. Two of these mission concepts, Large Ultraviolet/Optical/Infrared (LUVOIR) Surveyor mission and Habitable Exoplanet (HabEx) mission, address the need for a successor to the scientific legacy established by the Hubble Space Telescope (*HST*) and the exciting opportunity to search for life as a follow on to *Kepler*'s planet finding discoveries.

Mid-decadal studies have been ongoing that build on the work of the National Research Council's Decadal Survey for Astronomy and Astrophysics, *New Worlds, New Horizons* (*NWNH*).[1] These studies have consistently set forth technology development goals aimed at (1) enabling a future large aperture, UVOIR flagship mission as a successor to *HST* and (2) increasing the scientific reach of smaller missions. The Cosmic Origins Program Analysis Group (COPAG) is now evaluating and recommending technology investments towards these goals through Science Interest Groups (SIG)s, and the Exoplanet Program Analysis Group (ExoPAG) has also made this recommendation. In both of these scientific focus areas, single photon counting and ultra-low noise detectors are a priority. Furthermore, these recommendations set as a goal very large format (i.e., 100-Megapixel to Gigapixel), high quantum efficiency (QE), UV-sensitive detectors. Additionally, the Association of Universities for Research in Astronomy (AURA), convened a group to evaluate a next generation flagship large UVOIR space telescope. Their recommendation for the High Definition Space Telescope (HDST) includes ultralow noise and UV-sensitive detectors as highest priority technology, especially when produced cost-effectively and in high numbers. These new recommendations from AURA, COPAG-SIG, and Astro2010 reflect the new understanding of the scientific opportunities enabled by technological breakthroughs and large-scale detector fabrication.

Frontier astrophysical investigations are necessarily conducted at the limits of resolution, *étendue*, and sensitivity. A future 9-16-meter UV/optical telescope mission will require significant detector advances beyond HST, Galaxy Evolution Explorer (*GALEX*), and Far Ultraviolet Spectroscopic Explorer (*FUSE*) detector technologies, particularly in QE, spectral responsivity in the UV, resolution, area, pixel count, and intrinsically low detector dark current. Dramatically increasing the efficiency of the detectors could allow Explorer class or Probe class missions to perform Flagship mission science.

In this paper, we discuss our development of high performance silicon detectors that are widely applicable to various UV/optical and UV missions in different classes from suborbital to orbital missions in Explorer, Probe, and Flagship class missions. We give an overview of the deployment of these detectors and discuss their potential future applications.

The development and deployment of delta-doped and superlattice-doped (also called "2D-doped", as described in Sec. 3) silicon arrays is summarized pictorially in Fig. 1.1. Delta-doped arrays arose from efforts to address the surface passivation problems faced by the *HST* CCDs, and the first delta-doped CCD was reported in Hoenk et al. 1992.[2] This first device was a 100×100 pixel array and exhibited 100% internal QE; by the mid 1990's, we had demonstrated arrays as large as 1K×1K with improved stability and uniformity. Since then, JPL's delta-doped arrays have been demonstrated on wafer-scale devices independently evaluated and their performance has been verified by various groups, including the group led by the *Kepler* PI for precision photometry.[3] Additionally, delta-doped arrays of various formats have successfully flown on multiple suborbital flights, including sounding rockets with the University of Colorado, Johns Hopkins University,

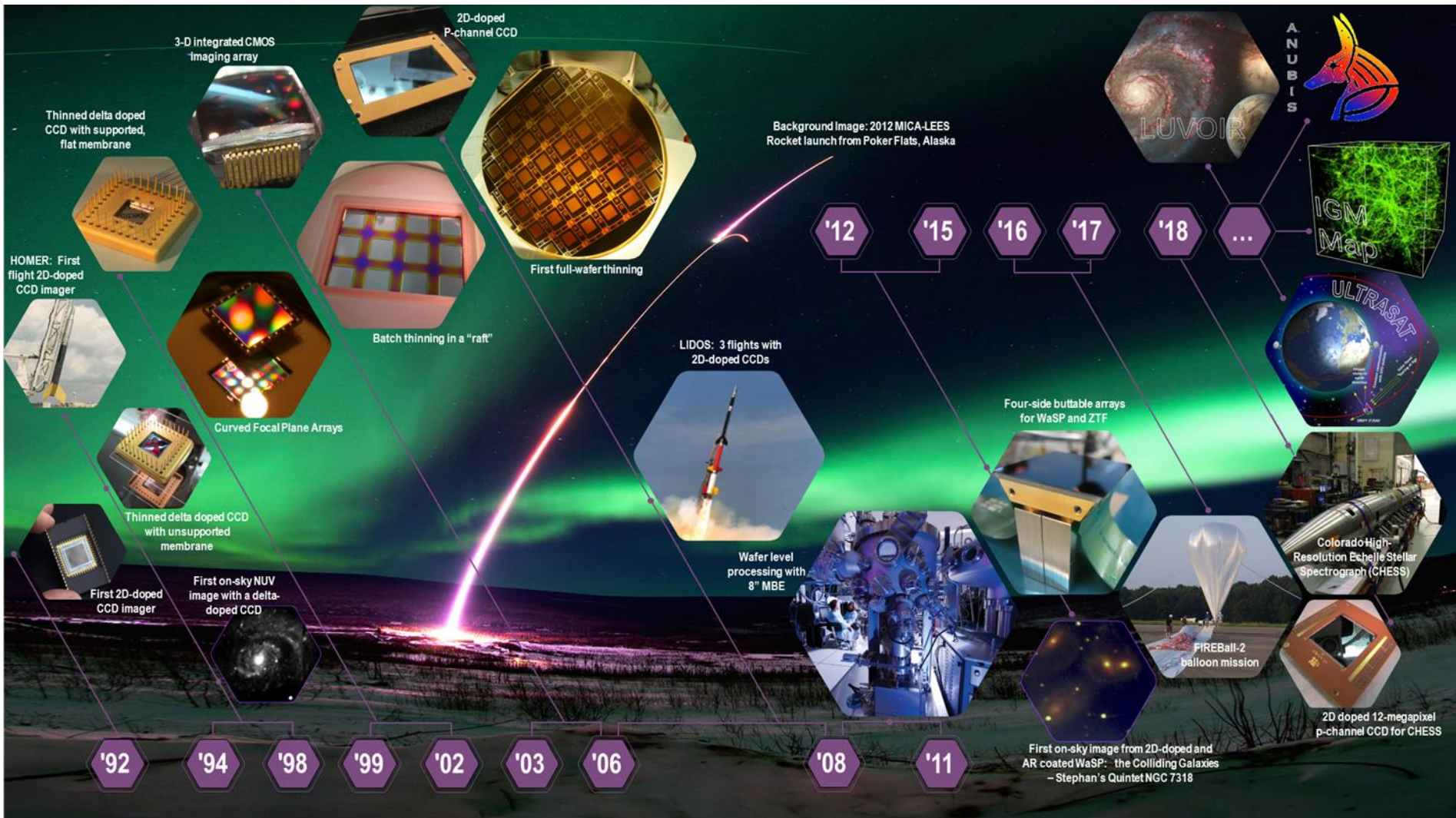

**Figure 1.1**. More than two decades of development and deployment have proven the capabilities and expanded the scale of delta-doped silicon arrays. This technology is well poised to make a significant impact on the capabilities of future Explorer-class and Flagship missions.

and Cornell/ Dartmouth/ Southwest Research Institute.

Delta-doped arrays have also been delivered for the Faint Intergalactic Redshifted Emission Balloon experiment (a stratospheric balloon payload) and the Colorado High-resolution Echelle Stellar Spectrograph (a rocket payload). Recent and ongoing ground-based observations include work at Palomar's Cosmic Web Imager and at Steward Observatory. Work completed during the past two decades have advanced the delta-doping technology such that it is perfectly poised to be incorporated into future Explorer-class, Probe-class, and Flagship missions.

## 2    Scientific Detectors for NASA Missions

*2.1  Silicon Detectors and Imaging Arrays*

Silicon imaging and detector arrays, especially charge coupled devices (CCDs) and complementary metal oxide semiconductor (CMOS) arrays, are ubiquitous in different fields of imaging applications. This is due in part to the steady advancement of silicon VLSI technology and the consumer market for imaging. Shortly after the invention of CCDs in 1969 at Bell Labs, extensive programs were established at JPL to advance early CCDs for imaging systems aboard NASA's Flagship mission *Galileo*[4] and its first great observatory, *HST*.[5] CMOS-based imaging started around the same time in the 1960's,[6] but it was not until 1990's that their development began in earnest due in part to fabrication advancement and to focused effort of CMOS-Active Pixel Sensor development at JPL and other CMOS images elsewhere.[7,8] CMOS imaging has had an enormous impact on consumer field and in recent years has become viable for scientific applications.

Through the years, the metrics of detector performance have progressed as VSLI technology has matured. Today, silicon imagers are commercially available and cost far less than the developmental detectors that NASA funded for *Galileo* and *HST*. The imager array size has gone from 800×800 (*Galileo*) to wafer-scale (e.g., 10K × 10K). At the same time, many other silicon detector technologies, including some that are capable of single-photon sensitivity, have been developed to meet the growing demand for performance application-specific detectors (Table 2.1). While each of these device types differ in their architectural details, there is much overlap in terms of device physics underlying spectral sensitivity and dark current. The processes described in this manuscript for surface passivation and UV sensitization (Sec. 3) are widely applicable to each of the devices described in Table 2.1, as well as other silicon-based arrays.

Table 2.1. Summary of silicon-based device architectures and application considerations for each.

| Detector | Description & Considerations |
|---|---|
| CCDs | First digital imaging arrays in space, CCDs have characteristically linear response, high uniformity and low noise. They are read out serially which requires multiple charge transfer steps. Workhorse for scientific imaging. Wafer-scale production of scientific-grade detectors is possible |
| CMOS | Similar in concept to CCDs but with parallel readout enabled by "amplifier per pixel" capability. Highly driven by commercial applications, low-noise CMOS imaging arrays are becoming available for scientific applications. Parallel readout with per pixel amplifier allows for fast/versatile readout. |
| High Purity Silicon Devices | CCDs made from high purity/high resistivity material. The entire silicon volume is depletable to 200-300 $\mu$m depths.[9] As such, NIR ($\lambda$~1 µm) photons can be detected efficiently and without fringing, giving improved red response. Thinning to epitaxial thicknesses (5-15 $\mu$m) not required. |
| EMCCD: Electron multiplying CCD | "CCDs with gain". An additional serial register, the "multiplication" or "gain" register. With a modest applied voltage (40-50 V) a gain of 1000s can be achieved. Essentially eliminates read noise. Single photon counting capability. Can be operated in "normal mode" or in "photon counting mode", effectively increasing the dynamic range of the detector |
| Hybrid Arrays | Detector circuitry and readout circuitry are fabricated in separate silicon wafers. These wafers are then physically/electrically interconnected; typically, with indium bump-bonding. Hybrid arrays combine high sensitivity and fast readout capabilities. Detector and readout may be optimized independently. |
| Low Noise CMOS | Also called "scientific" CMOS (sCMOS). Combines low-noise and robustness for space-based applications. |
| APD: Avalanche Photodiode | Semiconductor version of a photomultiplier tube. High voltages (hundreds of volts) across the devices results in avalanche multiplication of charge. |
| SPAD: Single Photon Counting Avalanche Diode | Reverse-bias PN avalanche diodes that operate in the Geiger mode to produce large current pulse; i.e., CMOS imager in which gain can be achieved within each pixel.[10–12] Single photon counting capability and fast timing resolution ideal for time-resolved measurements. |

## 2.2 Achieving High Quantum Efficiency and Broad Spectral Response in Silicon

The QE and spectral range of front-illuminated silicon detectors are limited by absorption and scattering by metals and polysilicon structures that form the essential circuitry on the front surface, regardless of the detector readout structure (e.g., CCD, CMOS, or APD). Very early in the history of silicon CCDs, it was realized that backside-illumination offers a convenient way to circumvent these limitations, potentially enabling broader spectral response and higher QE. However, new limitations and technological challenges were immediately created by back illumination. Early back-illuminated detectors were plagued by time-variable and nonuniform back-surface charging, which resulted in severe problems with low, nonuniform and unstable QE. The severe quantum efficiency hysteresis (QEH) in WF/PC-1 CCDs launched an intensive R&D effort at JPL to develop a stable back-illuminated CCD. The history of JPL's delta-doped CCD technology can be traced to the focused effort on solving the QEH problem .

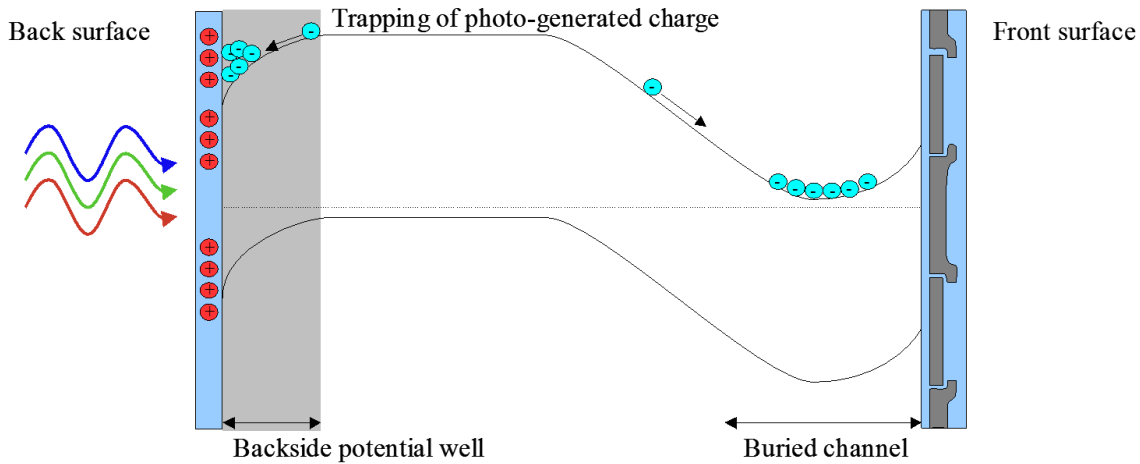

**Figure 2.1**. Schematic drawing of the electronic band structure in a back illuminated silicon imaging detector.[13]

Figure 2.1 shows a cross section schematic drawing of the electronic band structure of a back-illuminated silicon imaging detector. The front surface electrodes are shown schematically on the right, and an oxide formed on the back surface is shown schematically on the left. Light enters through this thin oxide generating free charge carriers in the conduction band. In order to be detected, the charge carriers must move by drift and diffusion into the buried channel near the front surface. However, defects at the Si-SiO$_2$ interface create electronic states that can trap and release electrons and holes, resulting in time-variable surface charge and electric fields near the silicon surface. These fields can trap photogenerated charge, causing time-variable (unstable) QE; this is the source of QEH. Surface charging affects device performance at all wavelengths, but the shallow absorption depth of UV photons renders UV detectors especially vulnerable to surface defects and associated problems with low and unstable QE. Surface defects are also responsible for elevated surface-generated dark current, which introduces noise and reduces the sensitivity and dynamic range of scientific detectors.

In order to achieve stable, high QE in back-illuminated silicon detectors, and to reduce surface dark current, the surface must be passivated. However, the methods for passivation are constrained by the detector itself. Most surface passivation processes require high temperatures that would damage or destroy the detector. Several approaches have been developed to passivate and stabilize back-illuminated silicon detectors, including:

1. Although not a surface passivation technique, the use of phosphorescent materials (e.g., coronene, lumogen) to down-convert UV photons to visible wavelengths was used as a "work around" to achieve UV sensitivity and improved QE.[14] This approach was used in the Wide Field Planetary Camera on *HST*.[15] Although phosphorescent materials help to reduce the magnitude of the problem, they must be used in conjunction with backside charging or surface doping in order to eliminate quantum efficiency hysteresis entirely.

2. Backside charging methods (e.g., UV flood, flashgate, and/or chemisorption charging), which introduce net negative charge in the oxide to overcome any positive charge trapped in defects at the Si-SiO$_2$ interface. The main challenge for backside charging methods is one of stability, as dissipation of charge over time can necessitate the use of methods (e.g., a UV flood) to recharge the surface.[16]

3. Surface doping techniques, which stabilize the detector by reducing the thickness of the surface depletion layer and create an internal barrier that suppresses surface dark current.

JPL's approach to surface passivation is based on a unique form of backside doping, which is described in detail in Sec. 3.

## 3 Delta-doped Silicon Detectors

*3.1 Post-Fabrication Processing of Silicon Detectors*

By using silicon-based detectors, the astrophysics, heliophysics, and planetary science communities are able to take advantage of the enormous investment made in the silicon industry. Even so, there are still significant challenges to achieving high quantum efficiency that spans the UV/optical/NIR wavelength range and high yield production of "science grade" devices. This section describes our development of post fabrication processing and delta doping and superlattice doping, which effectively and efficiently addresses these challenges and has been applied to multitude of silicon array architecture and a high quantity of wafers and devices.

The photosensitive volume of most silicon detectors comprises a thin layer of high purity epitaxial silicon that is grown on a thick, highly conductive silicon substrate. Highly conductive silicon makes a poor detector, as photogenerated electrons are lost to recombination before they can be detected. In order to achieve high QE in a back illuminated detector, the thick substrate must be removed to expose the thin layer of high purity silicon. Thinning (i.e., the substrate removal process) a silicon substrate that is hundreds of microns thick while leaving intact the 5-15 µm silicon detector proves to be extremely difficult. Whereas early detectors were relatively small, the challenges of thinning have grown with the size of detectors. Even after the substrate has been removed, two essential problems remain. First and most obviously, the silicon substrate performs an essential function of mechanical support. An unsupported silicon detector is fragile and frequently deformed by internal stresses into a shape that has been compared with a potato chip. Second, the silicon substrate also performs an essential electrical function. Being electrically conductive, the silicon substrate forms a stable, equipotential surface and a back-surface electrode that prevents photogenerated holes from accumulating on the back surface.

*3.2 Backside Thinning*

As previously discussed, backside illumination is required to achieve the best possible performance from silicon-based detectors. For backside-illuminated operation, all the excess silicon (i.e., the substrate) outside of the depletion region must be removed. For conventional CCD and CMOS, this requires thinning down to the epitaxial layer (5-20 μm); whereas for high purity, fully depletable devices and APDs, for example, thinning to 200-300 μm is sufficient.

Early detector thinning focused on "frame-thinning", in which the thinned region is supported by the surrounding thick frame of silicon from the original full-thickness device. The process for frame-thinning involves a sequence of liquid etches that terminate at the epitaxial interface between bulk, low-resistivity silicon and the high-resistivity silicon of the device's epitaxial layer.[17] With appropriate patterning techniques, the frame-thinning process can be carried out with a single detector, a raft of several detectors, or an entire wafer. Frame-thinning offered a relatively fast approach to producing back-illuminated, thinned arrays; however, the approach produces fragile silicon membranes—sometimes as thin as 5 μm (Fig. 3.1a). The frame-thinning approach is not a sustainable approach for scaling up to large imaging arrays or high throughput production of scientific imaging arrays.

A major advance came with the development of wafer-to-wafer direct-oxide bonding, in which two wafers are bound together *via* the intermolecular forces between two very flat surfaces. In the case of the detector wafer, a thick oxide is added, patterned and planarized. The device wafer front surface and that of a silicon support (or handle) wafer are activated by plasma etch, with the resulting surfaces each terminated with a labile bonding group.[18] The blank handle wafer serves (1) to protect the device wafer throughout the remaining processing steps and (2) as mechanical support for the thinned device wafer. Following the bonding step, the device wafer is thinned by grinding and chemical/mechanical polishing until the wafer thickness is reduced from hundreds of microns to tens of microns. The remainder of the substrate is removed by a selective thinning process, using a dopant-sensitive liquid etchant enables the use of the epilayer as an etch stop. A final polish produces a mirror finish surface.[19] Using this wafer-level thinning process, thickness variations on the order of 1-3 μm can be achieved across an 8-inch wafer (Fig. 3.1).

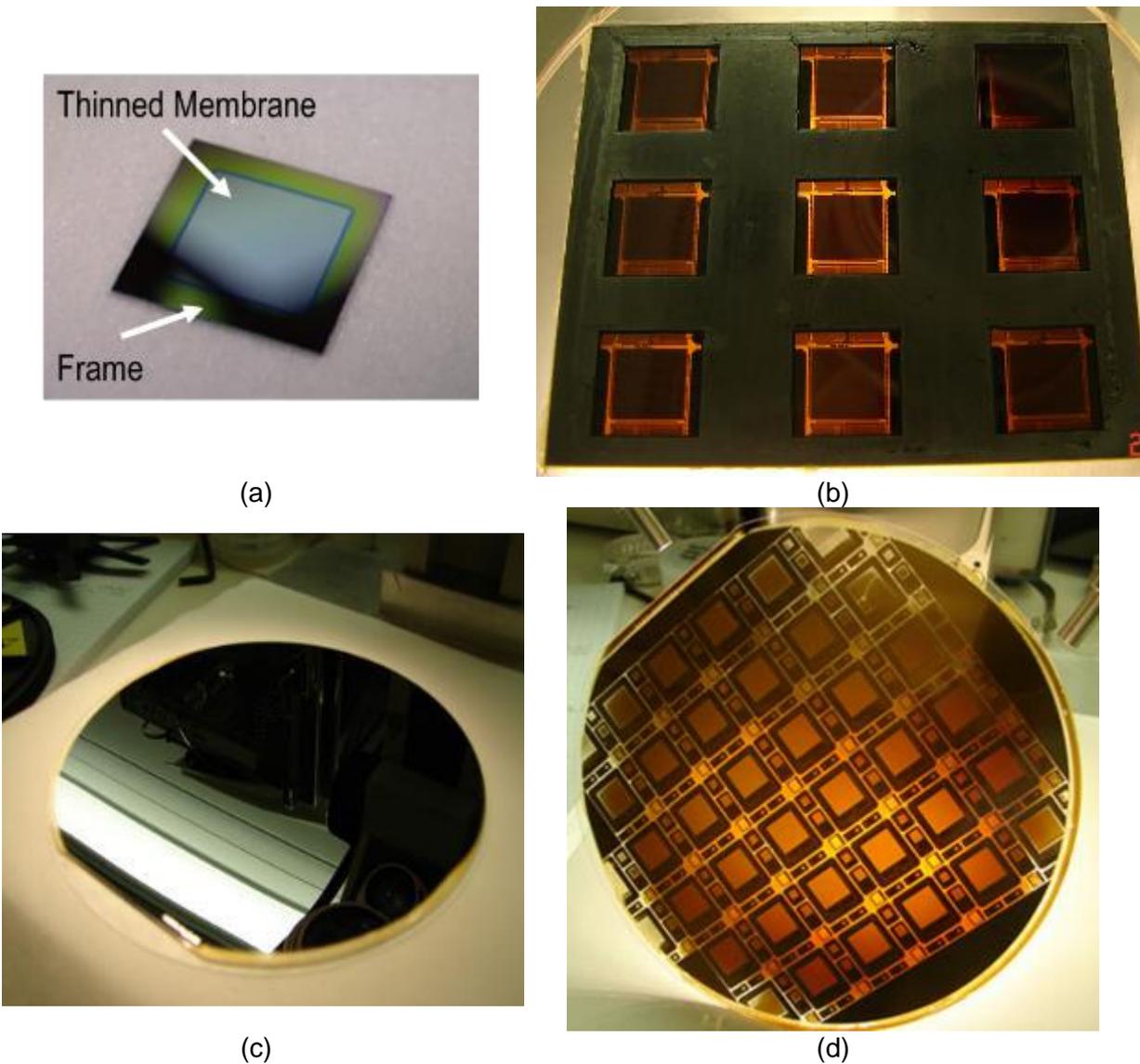

**Figure 3.1**. (a) Frame-thinned single die and (b) a frame-thinned "raft" showing batch processing nine die that were thinned at once. (c) The backside of a bonded device wafer during the thinning process. (d) The backside of the same wafer thinned to 5 μm. In this case, the wafer is supported on a transparent quartz substrate; thus the entire structure is transparent.

It is important to note that the process of physically reducing the thickness of the silicon will affect the detector's long wavelength response. Figure 3.2 shows the photon absorption length in silicon as well as the transmittance of silicon at varying thicknesses, including 10, 50, 200 and 500 μm; from the plots it is clear that silicon becomes increasingly transparent as the substrate is thinned, which will decrease red response. For cases in which high red response is needed, a fully depletable, high resistivity device should be used.

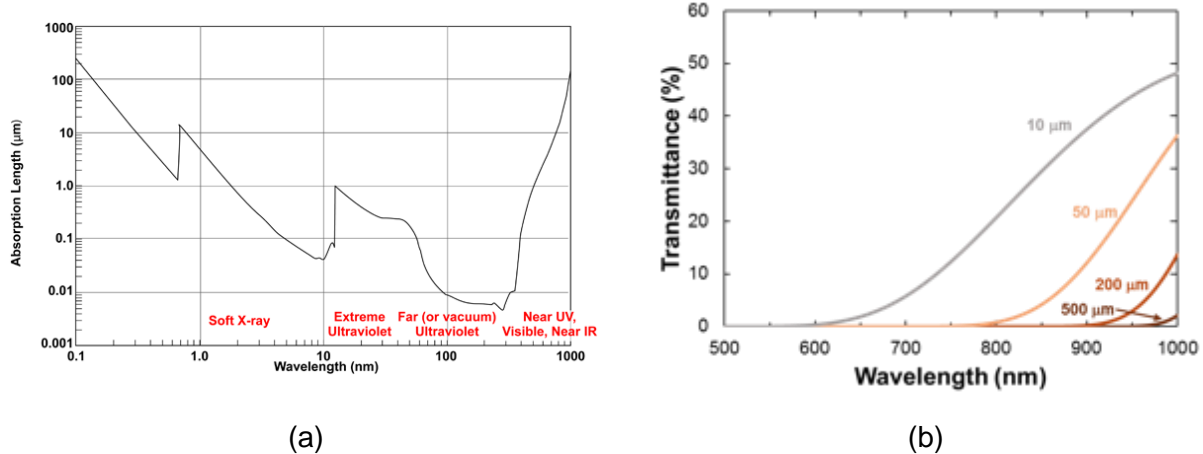

(a)                                            (b)

**Figure 3.2**. (a) Photon absorption length in silicon as a function of wavelength.[20] (b) Transmittance of silicon is plotted as a function of wavelength and for varying substrate thicknesses spanning 10-500 μm.

*3.3 Silicon Passivation by Two-Dimensional Doping*

As discussed in Sec. 2.2, back-illuminated detectors require surface passivation in order to achieve high, stable QE and reduce surface-generated dark current. Defects located at the Si/SiO$_2$ interface result in loss of efficiency and increased dark current. Because these effects are dynamic, they can make devices unstable, unreliable, and ultimately unusable. The goal of "surface passivation" is to create a stable, high efficiency detector. One approach is to reduce the defect density, and thereby reduce the charge density at the Si-SiO$_2$ interface. This approach played a key role in the history of the integrated circuits and silicon detectors. However, eliminating surface defects isn't enough for a detector for at least two reasons. First, a detector needs to collect photogenerated charge, so it is necessary to engineer the surface to provide a stable, built-in electric field. Second, imaging detectors designed for use in the space environment must be able to withstand exposure to ionizing radiation. Ionizing radiation creates defects at the surface that can severely degrade detector performance (see Defise et al. 1998 for data on EUV damage in ion-implanted CCDs)[21]. Of all of the approaches to surface passivation developed to date, none can match the sensitivity and stability of JPL's delta- and superlattice- doping.[13,22,23]

Delta doping and superlattice doping are two-dimensional (2D) doping techniques achieved by low temperature molecular beam epitaxy (MBE).[2,24–28] The use of low temperature MBE allows for control of the surface doping profile and band structure engineering with nearly atomic-scale precision. Conventional 3D doping processes produce random dopant distributions in the silicon lattice, and are constrained by the solid solubility limit to a maximum achievable dopant concentration. For example, the solid-solubility limit for boron is approximately $3.3 \times 10^{20}/cm^3$ at 1100 °C.[29] Beyond this limit, dopant activation and crystalline quality are degraded. 2D doping enables low temperature crystalline growth of silicon with dopant concentrations at least an order of magnitude greater than this limit, while still achieving nearly 100% activation and high crystalline quality. Atomic boron deposited by MBE on an atomically clean silicon surface forms a self-organized, 2D phase with surface densities up to 0.5 monolayer (approximately $3.4 \times 10^{14}$ B/$cm^2$ on a <100> silicon surface). Subsequent growth of epitaxial silicon encapsulates and stabilizes this 2D layer of boron in the silicon lattice.[30] The process is termed "delta doping" because the resulting dopant profile resembles the mathematical delta function. "Superlattice doping" refers to a doping profile in which more than one delta layer is incorporated into the MBE structure (Fig. 3.3).[26]

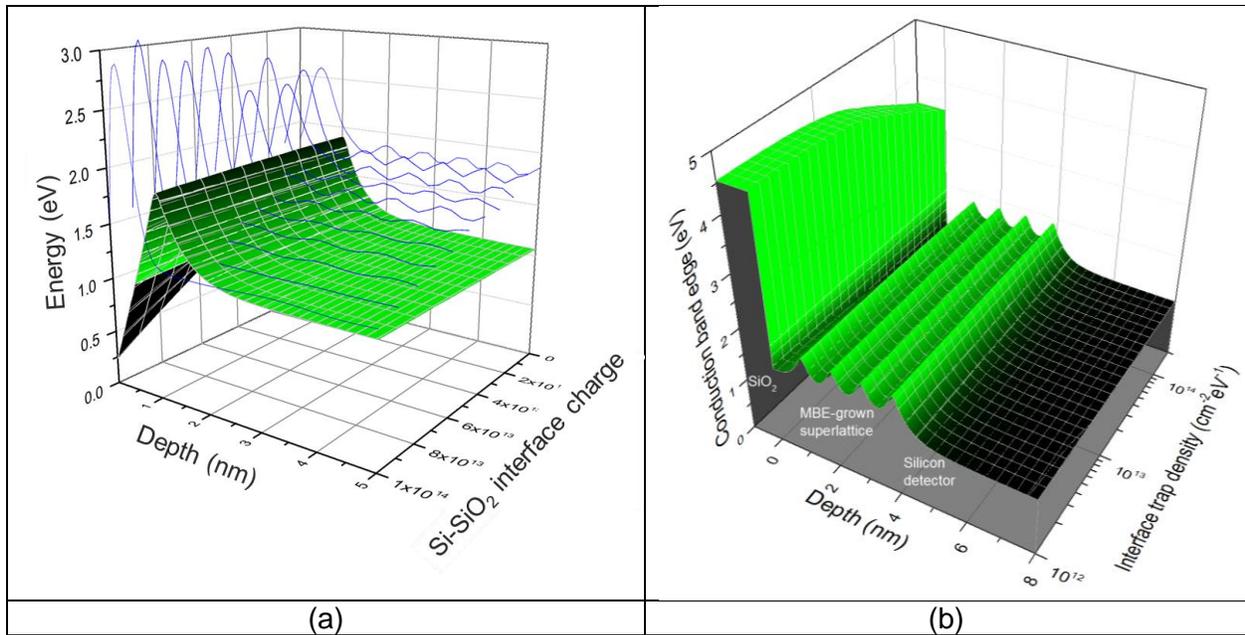

**Figure 3.3**. Plotted are the simulated band structure of (a) a delta-doped detector and (b) a superlattice-doped detector. Three-dimensional surface plots show the conduction band energy (z axis) as a function of depth from the Si-SiO2 interface (x axis), and interface trap density/charge (y axis). The width of the surface potential well determined by the MBE layer structure. The blue traces in (a) show representative electron wavefunctions; quantum confinement effects increase the ground state energy and prevent trapping of photogenerated electrons at the 2D-doped surface.

*3.4 Optical Coatings Prepared by Atomic Layer Deposition*

For low-light level applications, such as those described herein, it is critical to achieve the highest possible QE. JPL-invented 2D-doped silicon arrays exhibit 100% internal QE with their response following the reflection limit of silicon. Losses due to reflection can be significantly reduced through the use of antireflection (AR) coatings. Both hafnium oxide ($HfO_2$) and aluminum oxide ($Al_2O_3$) have been widely used as AR coatings with CCDs to improve visible QE (i.e., $\lambda > 400$ nm).[31,32] However, the rapidly varying optical properties of silicon in the UV wavelength range (Fig. 3.4) means there is no "one size fits all" AR coating for this challenging portion of the spectrum. Furthermore, in terms of material selection, one is limited to UV transmitting materials, including several metal oxides ($M_xO_y$) and metal fluorides ($MF_z$).

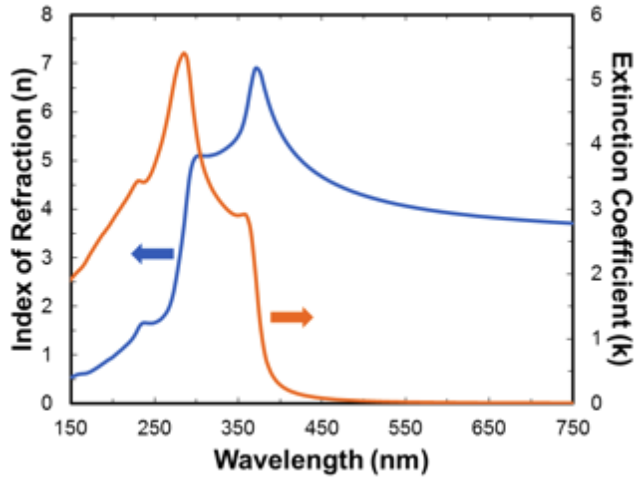

**Figure 3.4**. Optical properties of silicon, including the index of refraction (blue) and extinction coefficient (orange).

Additionally, the uniformity of scientific imaging detectors is imperative; thus, it is critical that AR coatings also be uniform and pin-hole free. Atomic layer deposition (ALD) is a thin-film technique that utilizes a series of alternating, self-limiting chemical reactions at the substrate surface. ALD has been widely used for the preparation of a variety of materials, including the UV transmitting materials already mentioned.[33] During ALD, films are deposited one monolayer at a time; thus it is possible to produce ultrathin, highly conformal, stoichiometric films with high density and excellent uniformity. Furthermore, ALD allows for smooth surface morphology and stacked films with well-defined interface characteristics, as shown in the atomic force microscopy (AFM) and transmission electron microscopy (TEM) images in Fig. 3.5.

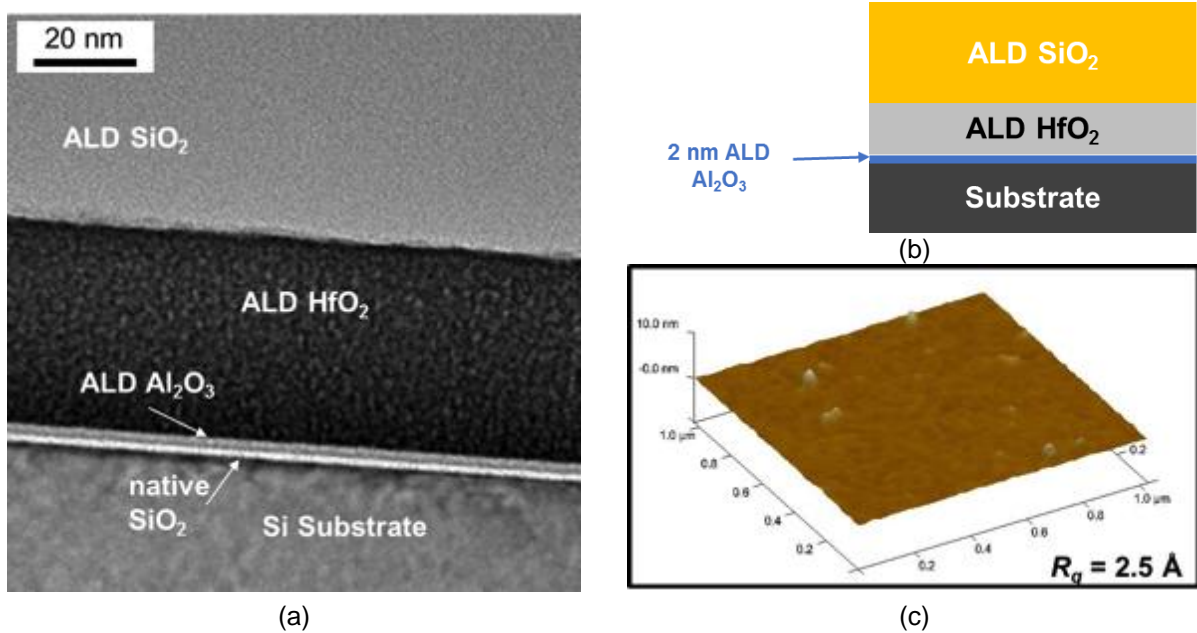

**Figure 3.5**. (a) TEM image of a stacked ALD films showing well defined interfaces and uniform layers. (b) Schematic diagram of the coating shown in (a) and (c). (c) AFM image showing a 1 μm² region of the coating shown in (b). The surface roughness is ~2.5 Å, similar to that of the uncoated surface.

The AR coatings described above offer a well-established approach to improving detector response, and can be tailored for both broadband and narrow band performance. However, there are certain cases in which suppression of out-of-band light is desired. For example, the detection of faint UV signals is often hindered by high visible background; the so-called "red-leak" masks the UV signal. In order to achieve good UV response together with red-leak suppression we turn to metal dielectric filters (MDFs) based on metal layers separated by transparent dielectric spacers. MDFs have long been used with transparent substrates for stand-alone filters, but only recently have we extended their use to silicon substrates for use as detector integrated filters. Our initial proof of concept studies based on Al/Al$_2$O$_3$ and Al/AlF$_3$ MDFs on bare silicon substrates (i.e., not device substrates) show excellent promise as UV band pass filters, suggesting that it is possible to achieve high in-band QE with out-of-band suppression on the order of $10^4$ (Fig. 3.6).

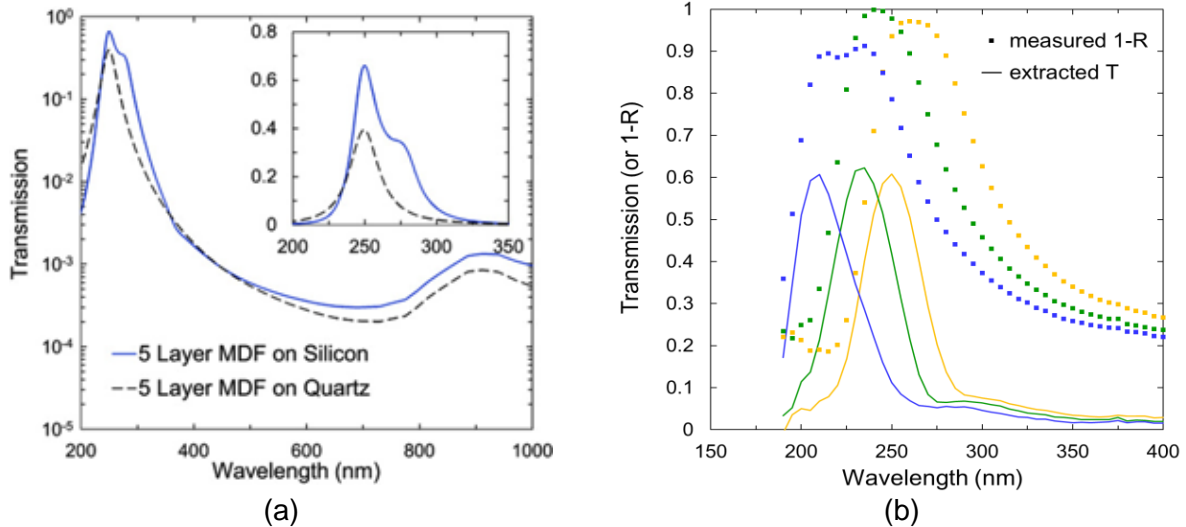

**Figure 3.6.** (a) Calculated performance of a five-layer MDF on silicon or quartz substrates. In band QE>60%, and out-of-band rejection approaches $10^4$.[34] (b) Prototype MDFs on Si using aluminum fluoride layers as an alternative to aluminum oxide with measured reflectance and projected transmittance based on ellipsometric characterization (not shown) for three designs optimized for 200–250 nm.

In addition to improved detectors, next-generation UV missions will require advancements in the realm of reflective optics, in particular improvements to the operating efficiency and environmental stability of standard Al mirrors at wavelengths shorter than the high-efficiency response of Al-MgF$_2$. Extending the short wavelength performance of conventional protected Al mirrors allows access to a rich spectral bandpass that has not been explored with high efficiency by previous NASA missions. The self-limiting nature of the ALD process makes it naturally scalable to large area substrates such as those envisioned for LUVOIR, and commercial systems have been demonstrated for the optical coating of meter-class substrates.[35] We refer the reader to our recent *JATIS* publication, Hennessy et al. 2016, which describes our work in this arena and provides a potential roadmap for future developments.[36]

The optical models presented in this manuscript were developed using the transfer matrix method and the TFCalc software package.[37] Optical constants for silicon, aluminum, and various metal oxides were taken from Palik and the Sopra database.[37,38] It is well known that a material's optical properties can vary depending on the deposition method; thus spectroscopic ellipsometry data collected using laboratory prepared samples (Horiba UVISEL 2; J.A. Woollam VUV-VASE) were also used in modeling where appropriate.

*3.5  2D Doping and End-to-end Processing of Different Device Architectures*

JPL's standard post-fabrication process as described above is summarized in Fig. 3.7.[19,39] Fully fabricated wafers are bonded to a handle wafer and back-thinned to the epi-layer. The wafer then undergoes several surface preparation steps to ready the surface for 2D doping and optical coating processes. Finally, the back surface is patterned and etched to reveal bond pads prior to dicing and packaging.

Our standard processing steps can be modified/customized to accommodate the various device architectures described in Sec. 2.1. For example, the high purity/high resistivity devices can be fully depleted to a thickness of 200-300 μm; eliminating the need for a handle wafer as well as the need to thin to the epitaxial layer. Without the handle wafer the front side electronics are exposed; however, it remains critical that the back surface is atomically clean and ready for epitaxial growth prior to MBE. Thus, surface preparation techniques that avoid chemistries that would attack the existing circuitry must be used. Finally, the backside pattern and etch (i.e., pad opening) is not required with high resistivity devices because the bond pads remain accessible from the front side. Similar modifications to the standard processes are made to accommodate APDs, hybridized arrays, and even front-side illuminated devices. Depending on the applications, we will implement our standard packaging or develop custom packaging as for example in the case of WaSP closely butted devices described in Sec. 4.2.4.

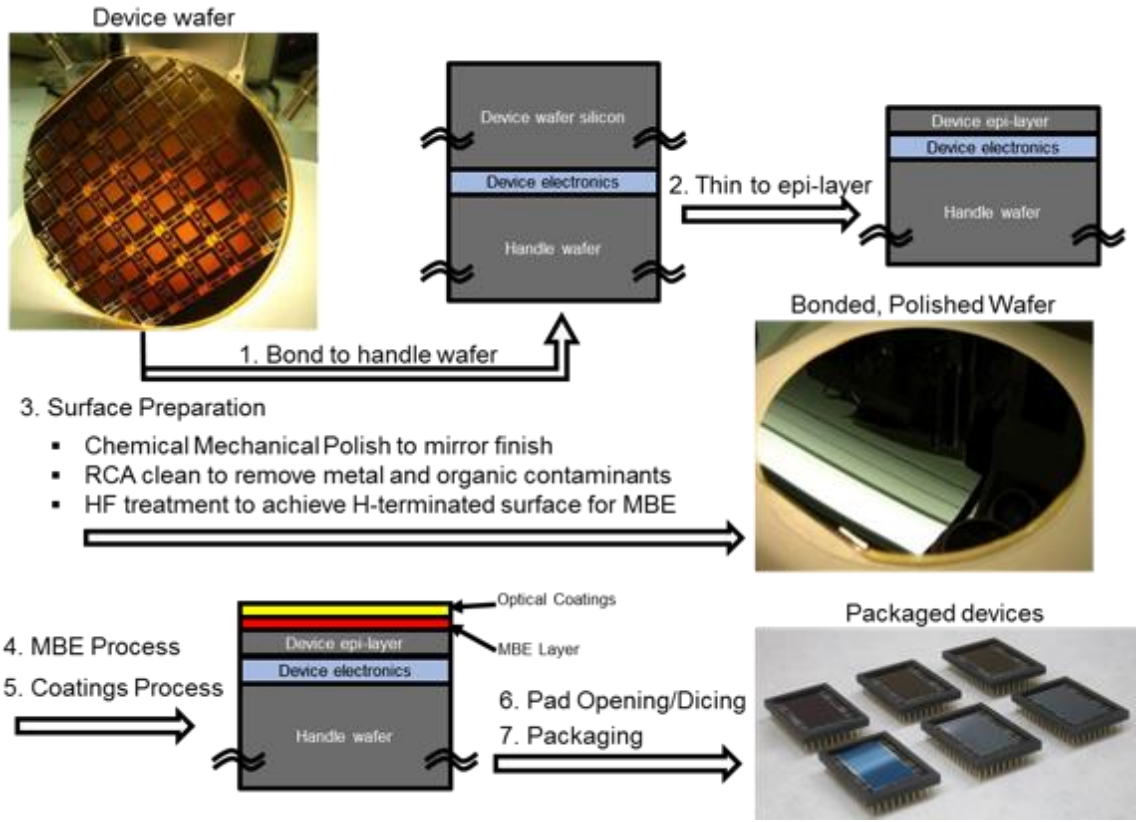

**Figure 3.7**. JPL's end-to-end post-fabrication process flow to produce high-performance UV/optical/NIR sensitized devices with 100% internal QE and tailorable external QE.[19]

*3.6 High Throughput Processing at JPL*

*Kepler*'s photometer is based on a 95-Megapixel focal plane, comprised of an array of 42 CCDs; thus bringing us into the era of space-based extremely large focal plane arrays.[40] Data from the *Kepler Space Telescope* has expanded our view of the universe by increasing our awareness of the commonality of exoplanets and confirming the existence of 21 (as of publication) "Earth-like" planets within the habitable zone. ESA's *GAIA*, launched in 2013, utilizes 106 CCDs (comprising a nearly 1-Gigapixel focal plane) in its mission to chart a three-dimensional map of the Milky Way and reveal the composition, formation and evolution of our galactic neighborhood.[41] Future missions will utilize even larger focal planes based on mosaicked arrays or monolithic devices.

State-of-the-art CCD fabrication facilities currently use six-inch wafers, and state-of-the-art CMOS imager foundries use eight-inch wafers. In order to enable full wafer scale production and batch processing of 2D-doped detectors, JPL developed infrastructure that includes acquiring a production-grade Veeco GEN200 Silicon MBE. This machine is equipped with a cluster tool with automated sample transfer system that moves wafer platens between chambers under computer control. Attached modules include a load-lock chamber and an ultra-high vacuum storage area with motorized elevators, enabling the loading and storage of up to eight wafer platens each for batch processing. Each ten-inch platen can be configured for single wafers up to eight inches in diameter or for multiple smaller wafers. The preparation chamber is differentially pumped, and equipped with a sample heater, gas inlet ports, and an RF source configured for implementing *in vacuo* surface preparation processes. The growth chamber has twelve effusion cell ports to accommodate multiple dopant materials, and dual e-beam sources enabling co-deposition of silicon and up to four additional source materials. The entire system is under computer control, enabling the development of automated high-throughput, multi-wafer processing.

Similarly, JPL's facilities include two ALD systems, both of which are capable of handling wafers up to eight inches in diameter. The Beneq TFS 200 ALD System is a high throughput, load-locked instrument. The growth chamber is equipped with inputs for six metal precursors, four reactive process gases and two thermal reactive sources; also included is a specialized input for "challenging" precursors (e.g., small quantities or low stability) for R&D purposes. Additionally, the Beneq is equipped with an exchangeable chamber and gas distribution unit; thus, chambers can be project or process specific, which will minimize contamination. The small volume of the growth chamber allows for highly precise temperature control, and the plasma source in the Beneq ALD is fully configurable, offering direct and remote plasma process capability. The Oxford OpAL

ALD System has a growth chamber equipped with inputs for up to three precursor compounds and six ports for reactive gases such as ammonia and hydrogen. In addition to the growth chamber, there is an integrated sample introduction glovebox that can hold and store multiple samples under dry nitrogen.

# 4 Examples of Deployment of 2D-doped Silicon Arrays

This section will describe the deployment of delta-doped and superlattice-doped arrays for sub-orbital and ground-based observations. We will also describe selected industrial applications and their relevance to NASA. Finally, we will describe several satellite missions currently under formulation that have baselined 2D-doped arrays.

## 4.1 Deployments to Space via Sub-Orbital Experiments

This section summarizes the suborbital flights of 2D-doped detectors processed in our laboratory, including delta-doped and superlattice-doped architectures. A summary of the devices to be discussed is given in Table 4.1.

Table 4.1. Summary of completed and planned deployments of 2D-doped devices in sub-orbital payloads.

| Mission | Device MFR | Device Type | Pixel Count | Pixel Size ($\mu$m) |
|---|---|---|---|---|
| HOMER | EG&G | N-channel CCD | 512×512 | 27×27 |
| LIDOS | SITe | N-channel CCD | 1100x330 | 24×24 |
| MICA-LEES | Loral Fairchild | N-channel CCD | 1024x1024 | 12×12 |
| CHESS | LBNL-DALSA | High Purity P-channel CCD | 3508×3512 | 10.5×10.5 |
| FIREBall-2 | e2v | EMCCD[†] | 1024×2048[*] | 13×13 |

[*]Commercially available CCD201-20 devices are typically operated in frame transfer mode with an image area of 1024×1024 active pixels.
[†]N-channel

### 4.1.1 HOMER: Conventional N-channel CCD

The High-altitude Ozone Measuring and Educational Rocket (HOMER; launched August 1996) was the third in a series of ozone-mapping rocket experiments managed by the Colorado Space Grant Consortium. Its primary objective was to determine the vertical distribution of the principle oxygen species related to ozone ($O_3$) chemistry—including $O_3$, molecular oxygen ($O_2$), atomic oxygen (O), and nitric oxide (NO)—along the rocket trajectory to obtain an understanding of the rate of $O_3$ production as a function of altitude. In the first of the previous missions (1992), the Colorado Student Ozone and Atmospheric Rocket used photometers centered at 2680 Å, 2959 Å, and 4515 Å. For the second mission (1994), the Cooperative Student High Altitude Rocket Payload added a spectrometer with a 512×512 pixel CCD detector. For the third flight, HOMER's ultraviolet imaging spectrometer made measurements of the Earth's limb in the far UV (215-265 nm) using a delta-doped CCD as the detector (Fig. 4.1). This early demonstration of the advanced UV sensitivity provided by delta doping successfully resulted in the measurement of the ozone column-concentration as a function of altitude. The delta-doped CCD was key to the success of the mission by its ability to detect and spatially register the faint UV signal from atmospheric constituents.[42]

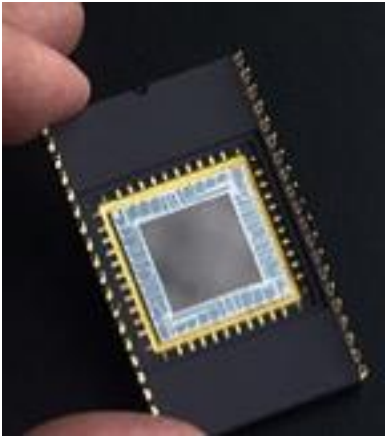
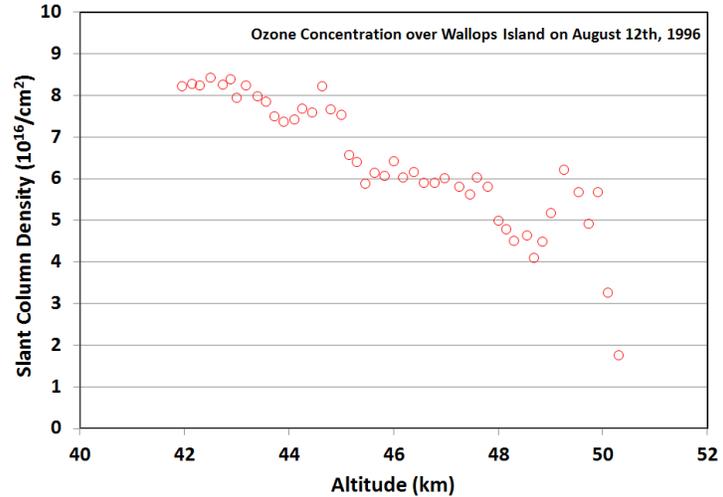
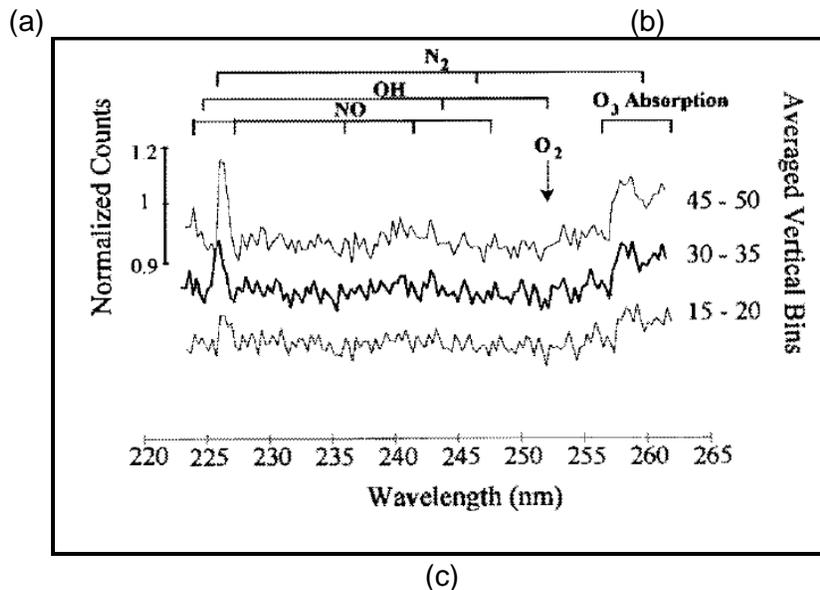

(a)                      (b)

(c)

**Figure 4.1**. (a) The HOMER detector, a 2D-doped 0.25-megapixel array. (b) Ozone concentration is shown as a function of altitude based on data returned from the HOMER mission. (c) UV Airglow spectra of the Earth's Limb from HOMER.[42]

### 4.1.2 LIDOS: Conventional N-channel CCD

Two bare (i.e., no AR coating), rectangular (1100×330) delta-doped SITe CCDs were delivered to Johns Hopkins University (JHU) for technology demonstration in the Long-slit Imaging Dual-Order Spectrograph (LIDOS). The delta-doped CCD was integrated into the LIDOS payload[43] and launched successfully aboard NASA sounding rocket missions 36.208 and 36.243.[44,45] Fixed exposure time observations were made of the UV-bright illuminating stars of the Trifid and Orion nebulae on the two flights. The detector is shown in Fig. 4.2a. The spectrum of $^1\theta$ Ori C (HD 37022) from the Orion nebulae mission (36.243) is shown in Fig. 4.2b. The presence of a non-black hydrogen absorption profile in the HI λ1216 transition, which is known to be highly saturated, suggests the presence scattered visible light photons (red-leak) in the spectrograph. A microchannel plate detector (MCP) was also included in the payload and exhibited a perfectly black HI λ1216 profile for this object. A windowless lamp, which produces far-UV bremsstrahlung continuum and line emission from electron impact on a tungsten target and residual vacuum gases, was used to illuminate the long-slit of the spectrograph to acquire simultaneous spectro/spatial flat fields of the two detectors. The comparison is shown in Fig. 4.2c.

Based on these side-by-side comparison data, it was concluded that "[an] ideal detector would combine the imaging performance and linearity of the CCD with the low background and photon counting ability of the MCP".[44] This work coincided well with the maturation of EMCCD technology and prompted our continued work towards the 2D-doped EMCCD, as discussed in further detail in Nikzad et al. 2016 and Sec. 4.1.5.[19] It should be pointed out that in most astronomical spectroscopic applications, the wavelength selectivity of the spectrometer and stand-alone filter combination removes the red leak and enables the use of silicon detectors. However, the final hurdle in use of silicon as solar-blind far UV (FUV) detectors is currently being addressed in in ongoing work at JPL as discussed in Sec. 3.3 and 4.3.1 and Hennessy et al. 2015.[34]

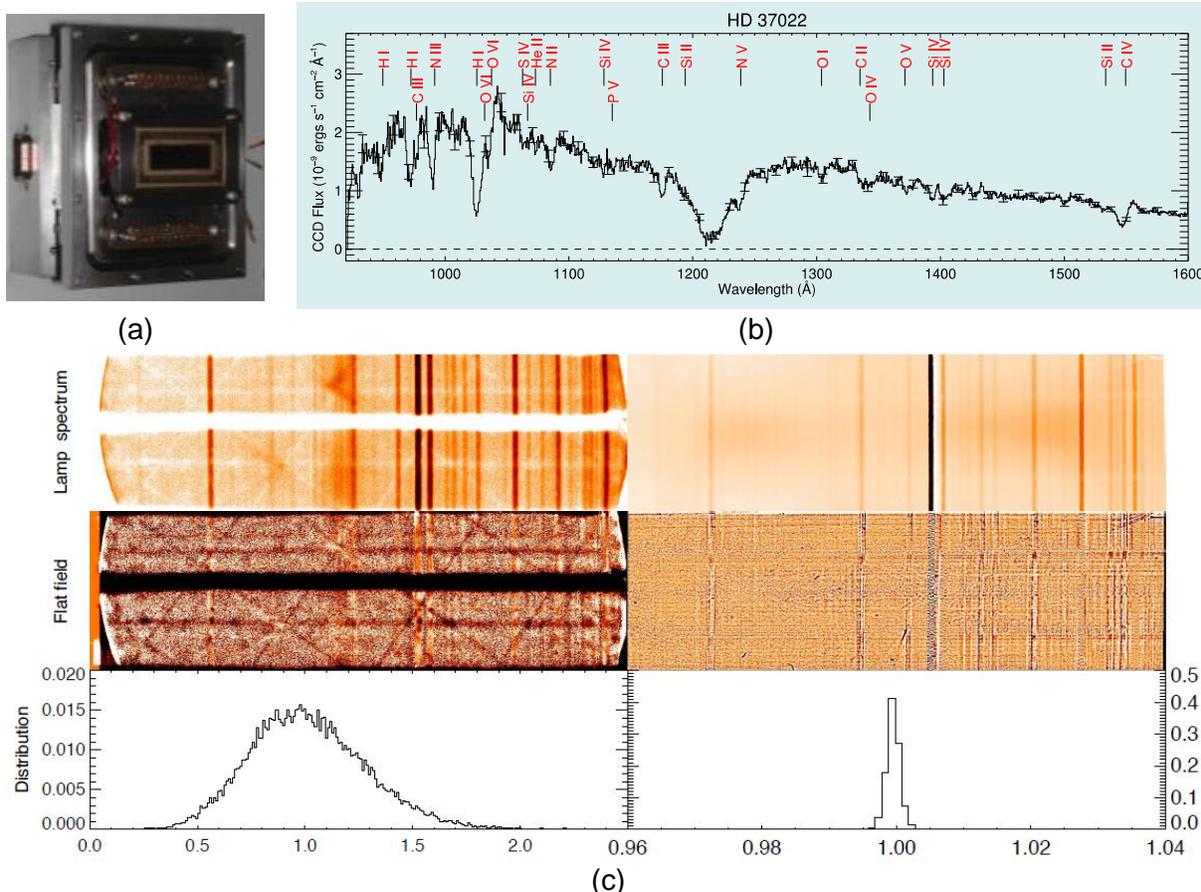

**Figure 4.2.** (a) Photograph of) the delta-doped CCD in the payload. (b) Flight data from NASA/JHU 36.243 UG.[44] Far-UV spectra of the central OB star cluster of the Orion Nebula (θ[1] Ori C) was obtained with the LIDOS delta-doped array. These data are used to constrain the properties of dust and molecules in photo-dissociation regions (see, e.g., France et al. 2005[46]). (c) Spectrum of the electron-impact lamp mounted in front of the entrance aperture of the spectrograph and post-processing flat fields for the MCP (left) and CCD (right) on board LIDOS. Histograms of the flat field variation are included at the bottom.

*4.1.3 MICA-LEES: Conventional N-channel CCD*

The Magnetosphere-Ionosphere Coupling in the Alfven Resonator (MICA) heliophysics sounding rocket mission incorporated a delta-doped array as the science detector in its Low Energy Electron Spectrometer (LEES). The objective of this experiment was to investigate the role of active ionospheric feedback in the development of large amplitude/small-scale electromagnetic waves and density depletions in the low altitude (<400 km), downward current, auroral ionosphere—a critical component in understanding magnetosphere-ionosphere coupling.

Delta-doped detectors allow direct detection of low-energy electrons without the need for high voltages power supplies and can be used at lower altitudes than possible with conventional detectors (e.g., MCPs). For this application, we operated the imaging detector in "photodiode mode", in which all of the output signal is added and read out as a current via a low noise circuit, digitized, and stored. MICA had a successful launch with a functional LEES. The LEES experiment served the purpose of demonstrating the successful flight and operation of a delta-doped particle detector; see Fig. 4.3.

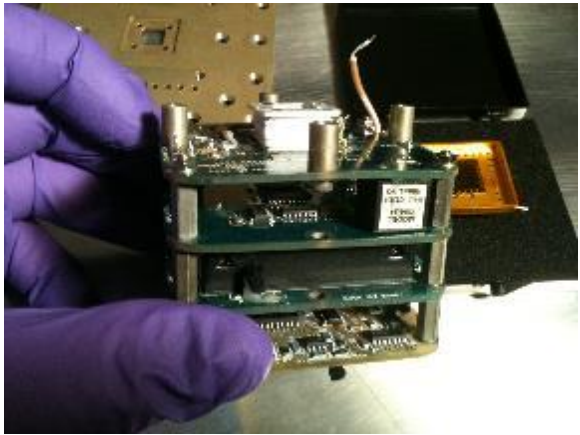
(a)

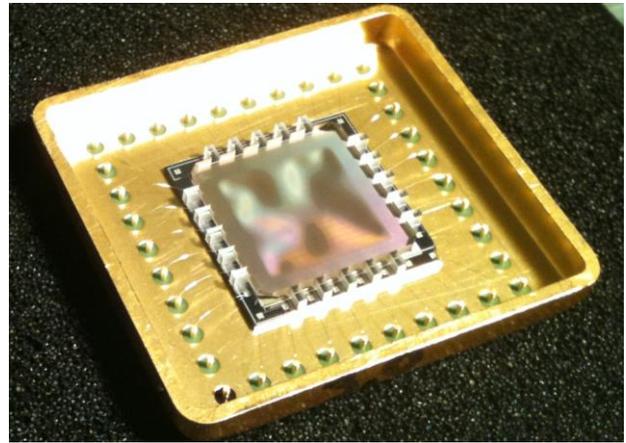
(b)

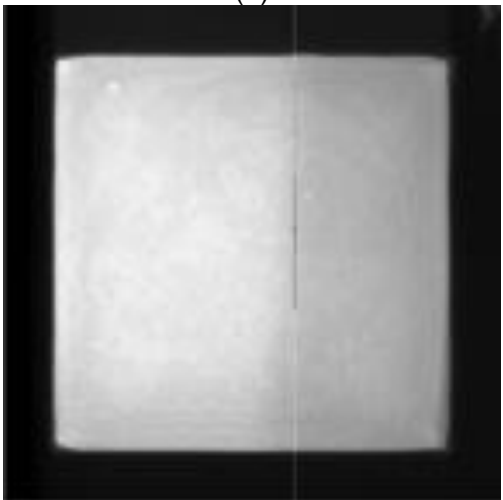
(c)

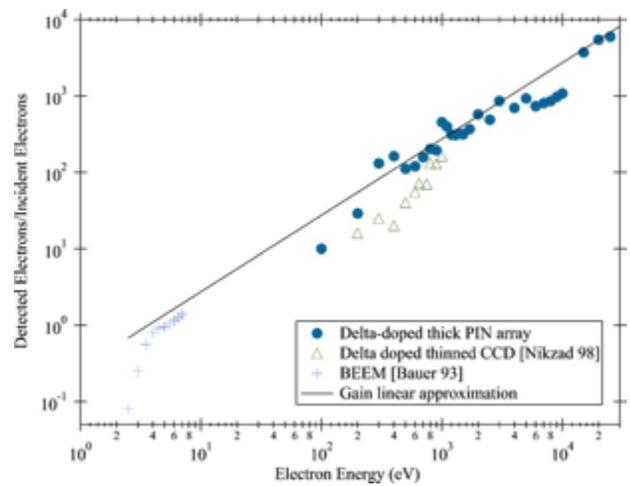
(d)

**Figure 4.3.** (a) MICA-LEES instrument and (b) photograph of the delta-doped CCD integrated into LEES. This device was produced early on with an unsupported, frame-thinned membrane. (c) A "flat field" image taken with the device in response to protons and (d) a typical electron response of device.[47]

*4.1.4 CHESS: High Purity P-channel CCD*

The Colorado High-resolution Echelle Stellar Spectrograph (CHESS)[48,49] is a rocket-borne far-ultraviolet spectrograph that serves as a pathfinder instrument and technology testbed for high-resolution spectrographs for future NASA astrophysics missions (e.g., LUVOIR[50]). The CHESS rocket experiment is designed to quantify the composition and physical state of the interstellar medium (ISM), specifically the interface regions between translucent clouds and the ambient diffuse ISM; quantifying the temperature, composition, and kinematics of nearby interstellar clouds. The local ISM provides an opportunity to study general ISM phenomena up close and in three dimensions, including interactions of different phases of the ISM, cloud collisions, cloud evolution, ionization structure, thermal balance, turbulent motions, etc. (see review by Redfield et al. 2006[51]). The CHESS instrument is an objective echelle spectrograph operating at f/12.4 and resolving power of R≈120,000 over a bandpass of 100-160 nm. CHESS has flown successfully in 2014 and 2016 using a cross-strip anode MCP detector system; a description of the flight performance of the CHESS instrument is provided in Hoadley et al. 2016.[52]

A future flight of CHESS will incorporate an echelle grating fabricated using advanced electron-beam etching techniques developed at Pennsylvania State University, and the payload will be further reconfigured to replace the MCP detector with a delta-doped, SNAP CCD—a 3508×3512 pixel, 10.5-μm square pixel format, high-resistivity, p-channel CCD designed by Lawrence Berkeley National Laboratory (LBNL) (Fig. 4.4).[9] The devices are 200 μm thick, back-illuminated, and packaged in a picture frame; because the CHESS spectral range extends to 100 nm, the detector will be flown without an AR coating.

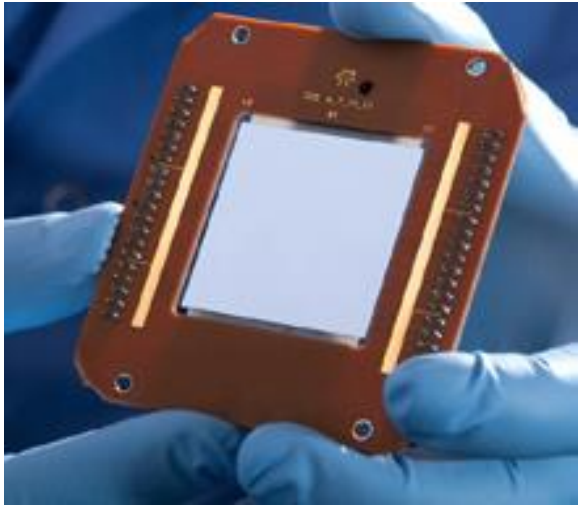 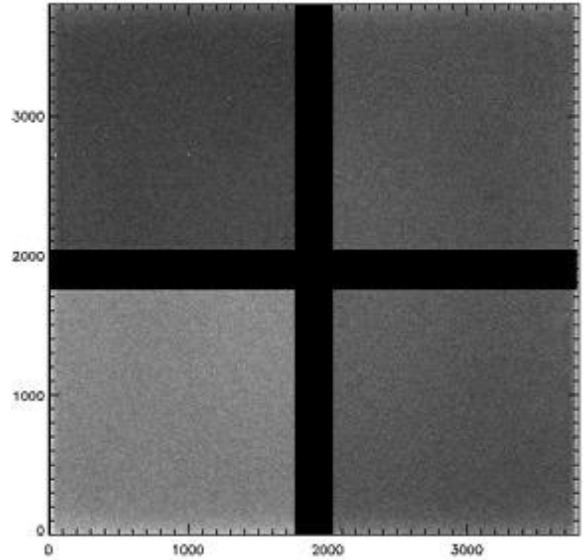

(a) (b)

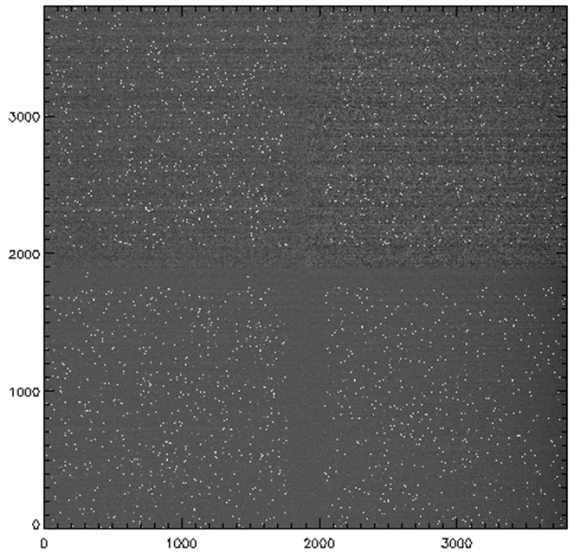 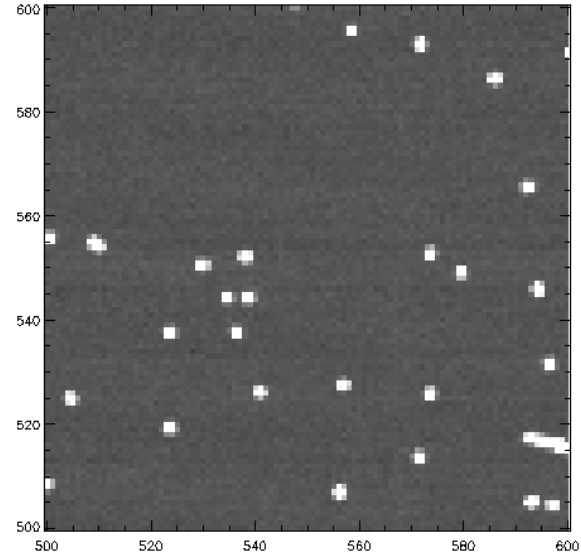

(c) (d)

**Figure 4.4.** (a) A 2D-doped, p-channel (SNAP) CCD that will fly on a future iteration of CHESS. (b) A 500 nm flat field exposure (5 s, -140 °C) with a 2D-doped SNAP CCD shows good uniformity. (c) X-ray measurements (20 s, -140 °C) shows excellent charge transfer efficiency (CTE). (d) Shows a 100 × 100-pixel X-ray image from the same device. CTE for this device is ≥0.999998 in all quadrants.

*4.1.5 FIREBall-2: EMCCD*

The Faint Intergalactic-medium Redshifted Emission Balloon (FIREBall-2) experiment is a balloon-born multi-object UV spectrograph designed to observe faint line emission from the circumgalactic media (CGM) of low redshift galaxies (z~0.7). A previous version of the FIREBall mission has flown twice, in 2006 and 2009.[53,54] FIREBall-2 will have improvements to the spectrograph that significantly increase the field of view, throughput, and number of targets per observation. These improvements will yield a factor of 30 increase in overall sensitivity allowing for multiple detections of the CGM in emission for the first time at UV wavelengths.

As a balloon-borne experiment operating in the upper atmosphere, FIREBall-2's UV operational range is limited to the stratospheric balloon window spanning ~195-225 nm. Its detector is a 2D-doped and AR-coated EMCCD (e2v CCD201-20). The commercially available CCD201-20 is operated in frame transfer mode with a total imaging area of 1024×1024; however, for FIREBall-2, the customized device operates in full frame mode with an imaging area of 1024×2048. Several AR coatings were considered for FIREBall-2, including single and multiple layer films. While multilayer AR coatings provide high QE with a narrow peak, variations in film thicknesses will result in a peak shift.[55] In the case of FIREBall-2, which is naturally constrained to a very narrow bandpass we consider AR coatings with five or fewer layers with a broader bandpass that encompasses the stratospheric window. Test coatings were prepared on test samples and functional test devices; test samples were evaluated in terms of their optical performance and related properties (e.g., surface and interface roughness) prior to moving to test devices. Transmittance data (100-R, %) from several of the FIREBall-2 coating development samples based on $Al_2O_3$ and $SiO_2$ are presented in Fig. 4.5, along with the predicted performance based on spectroscopic models. The single layer AR coating (orange data points in Fig. 4.5) was applied to a FIREBall-2 device, which was incorporated with the spectrograph. Bench-top tests with a zinc source show that the device has good sensitivity in the 200-215 nm range, as expected. Additionally, read noise and dark current data show that the 2D-doping process is compatible with the EMCCD platform.[56,57]

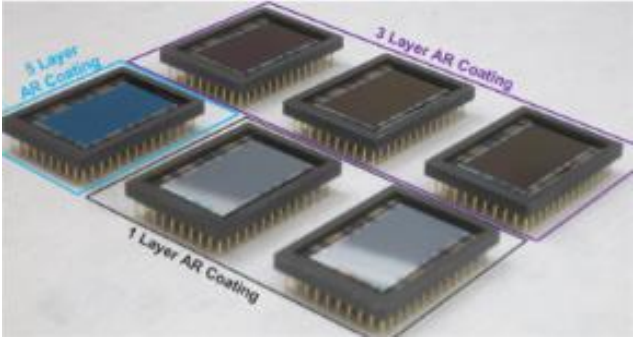

(a)

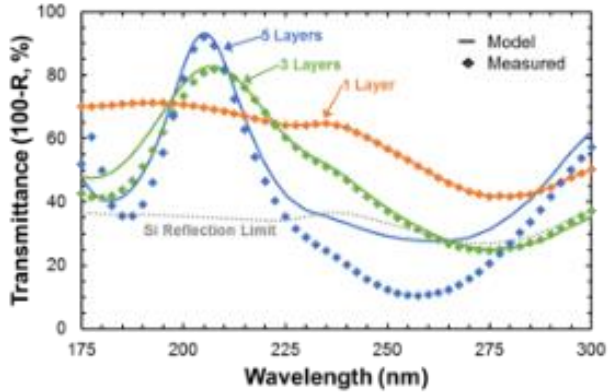

(b)

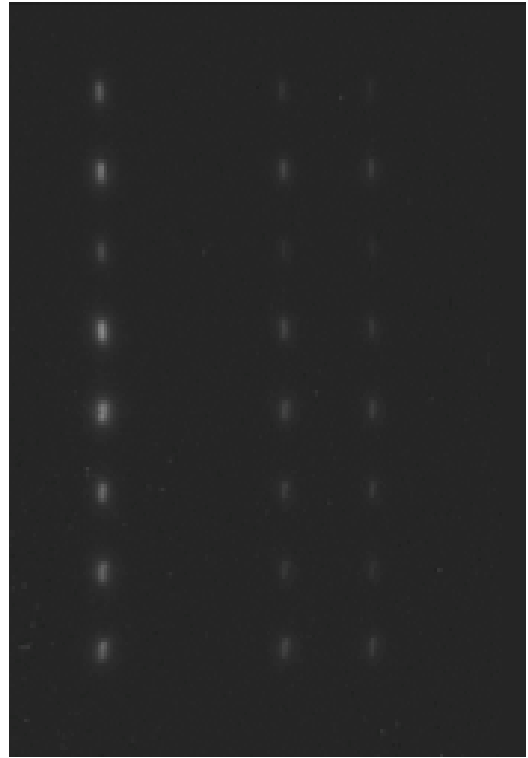

(c)

**Figure 4.5**. (a) Photograph of several 2D-doped CCD201 test devices for the FIREBall-2 project. These devices feature one-, three-, and five-layer AR coatings, distinguishable by their respective hues and labeling. (b) Model and measured transmittance (100-R) for FIREBall-2 test samples (i.e., not test devices). Also shown is the reflection limit for bare silicon (dotted line). Both the models and transmittance are calculated based on reflectance (R), and do not account for absorption within the film. (c) Zn line response (214, 206, 203 nm, left to right) on an optimized 2D doped EMCCD in the FIREBall spectrograph.

## *4.2 Ground-based Observations*

This section summarizes the use of 2D-doped detectors at ground-based observatories; Table 4.2 provides a brief overview of the detectors that will be presented.

**Table 4.2**. Summary of completed and planned deployments of 2D-doped devices at ground-based observatories

| Observatory | Device MFR | Device Type | Active Pixels | Pixel Size (μm) |
|---|---|---|---|---|
| Palomar (I) | EG&G Reticon | N-channel CCD | 512×512 | 30×30 |
| Lick | LBNL | High Purity P-channel CCD | 2048×4096 | 2048×4096 |
| Steward | LBNL-DALSA | High Purity P-channel CCD | 3508×3512 | 10.5×10.5 |
| Palomar (II) | STA | High Purity N-channel CCD | 2064×2064 | 15×15 |
| Palomar (III) | e2v | EMCCD[†] | 1024×2048* | 13×13 |

*Commercially available CCD201-20 devices are typically operated in frame transfer mode with an image area of 1024×1024 active pixels.
[†]N-channel

*4.2.1  Palomar (I): Conventional N-channel CCD*

The first on-sky images of delta doped CCDs provided rapid qualitative feedback from astronomer Jim McCarthy and team. A 512×512, 30-micron pixel, bare (i.e., no AR coating) delta-doped EG&G Reticon CCD was taken to Palomar Observatory for quick-look on-sky imaging in the near ultraviolet. Figure 4.6 shows two images of the same galaxy acquired in the (a) near UV (NUV) taken by the delta-doped CCD and (b) the visible image from the Atlas. While there are some overlapping features, the UV image offers a more detailed view of the center of the galaxy and provides information about the hot surrounding areas. Specifically, in the UV the younger hotter stars that form on the arm of the galaxy are easily distinguishable.

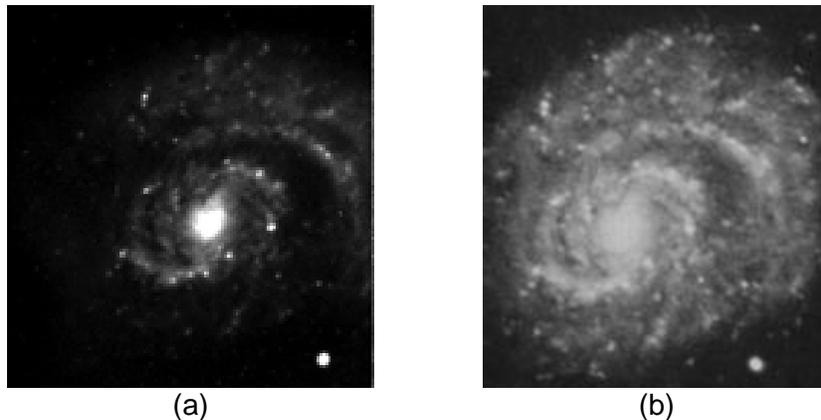

(a)                                             (b)

**Figure 4.6.** First on-sky images with a 512×512 delta-doped EG&G Reticon CCD in NUV. (a) Spiral galaxy 6137 imaged in the NUV (300 nm). (b) The same galaxy at imaged at visible wavelength (from astronomy Atlas).

*4.2.2  Lick: High Purity P-Channel CCD*

A delta-doped, p-channel, 8-megapixel (2048×4096) LBNL CCD with high broadband efficiency and high resolution (Fig. 4.7) was used for on-sky tests at Lick Observatory.[58] While this device sustained some damage during the integration in the dewar at the observatory, it was used to produce on sky images such as those shown in Fig. 4.7.

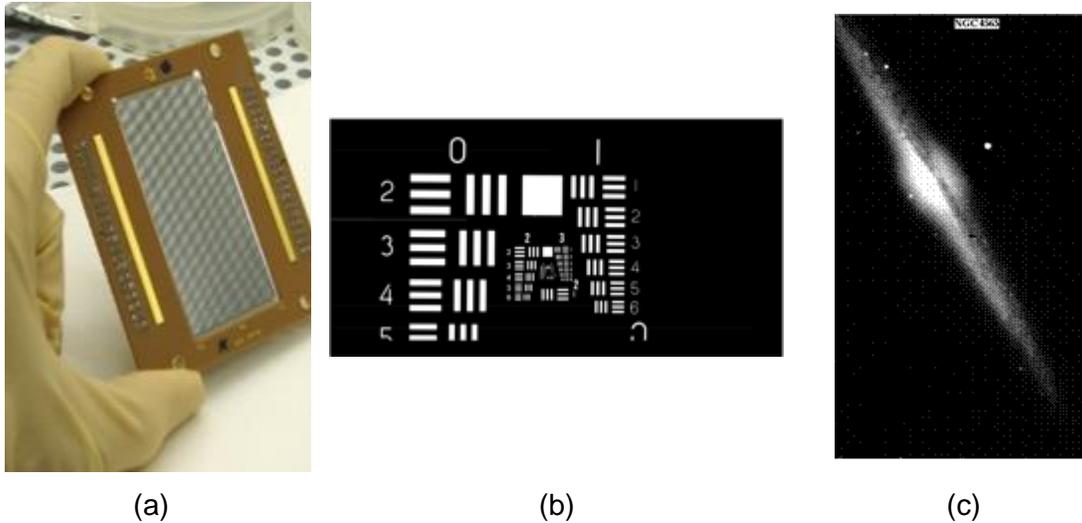

       (a)                         (b)                    (c)

**Figure 4.7.** (a) Photograph of an 8-megapixel, 2D-doped, p-channel LBNL CCD. (b) A high resolution pattern image with the same device. (c) First on-sky images with a 2D-doped, p-channel LBNL CCD (2048×4096).

*4.2.3 Steward: High Purity P-Channel CCD*

The CHESS flight prototype detector—a delta-doped, p-channel 3508×3512, 10.5-µm pixel CCD (SNAP, LBNL)—was incorporated in the dewar at Arizona State University (ASU) and used for on-sky observation and quantitative analysis of the QE, etc. Figure 4.8 shows the images of astronomical objects during the expedition.

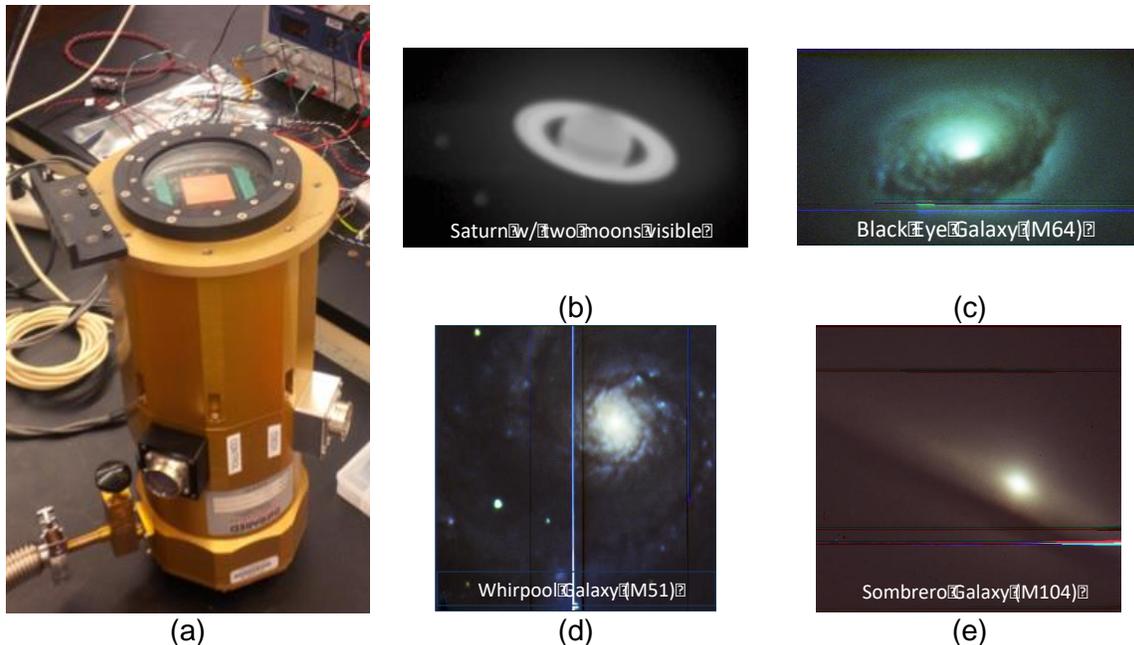

**Figure 4.8**. (a) An AR-coated version of the CHESS prototype detector installed at Arizona State University and subsequently mounted on the back of the 61" Kuiper telescope at Steward Observatory Steward Observatory. (b-e) First on-sky images with a 2D-doped and AR-coated SNAP CCD.

*4.2.4 Palomar (II): High Resistivity N-channel CCD*

The Wafer-Scale camera for Prime focus (WaSP), a collaboration between Caltech Optical Observatories and JPL, is the new prime focus imager for the 200-inch telescope at Palomar Observatory, providing multi-band color imaging from 320-1000 nm. WaSP was developed to replace the older Large Format Camera with a single high performance focal plane. Additionally, WaSP also serves as a technology pathfinder for the Zwicky Transient Facility (ZTF) on the 48-inch Oschin Telescope at Palomar, which will conduct high cadence sky-surveys for transient astronomical objects. In order to achieve the highest image quality, the instrument needs dedicated guide and focus CCDs to provide real-time feedback to the telescope tracking systems and focus controls. The guide CCDs, in particular, need to have as high sensitivity as possible—especially in the U-band as they observe through the science filter. Additionally, due to the tight space constraints in the WaSP and ZTF dewars, the guide and focus CCDs needed to be positioned as close as possible to the main science CCD within each system.

In response to these requirements of high broadband sensitivity and close-proximity device mounting, JPL developed detectors based on the 2064×2064, 15-µm pixel format back-illuminated STA3600 CCD in a custom four-side buttable packaging. The STA3600 is a high resistivity, fully-depleted CCD developed by Semiconductor Technology Associates, Inc. (STA). The increased silicon thickness provides enhanced response at the red end of the visible spectrum, where the sensitivity of conventional thinned CCDs drops off rapidly. These devices were modified for UV-enhanced sensitivity using JPL's 2D-doping and AR coating technologies. The custom CCD package (shown in Fig. 4.9) reduces the detector footprint to the physical size of the detector die, with no wire-bonds or detector packaging projecting outside that area. The packaging requirements also called for meeting tight tolerances on the final package thickness, to avoid shimming (on a per detector basis) in the dewar to achieve focus. All CCDs (guide, focus & science) have been installed on a flexure mount in the correct focal plane positions. Package thickness is repeatable to within 20 µm across multiple devices.

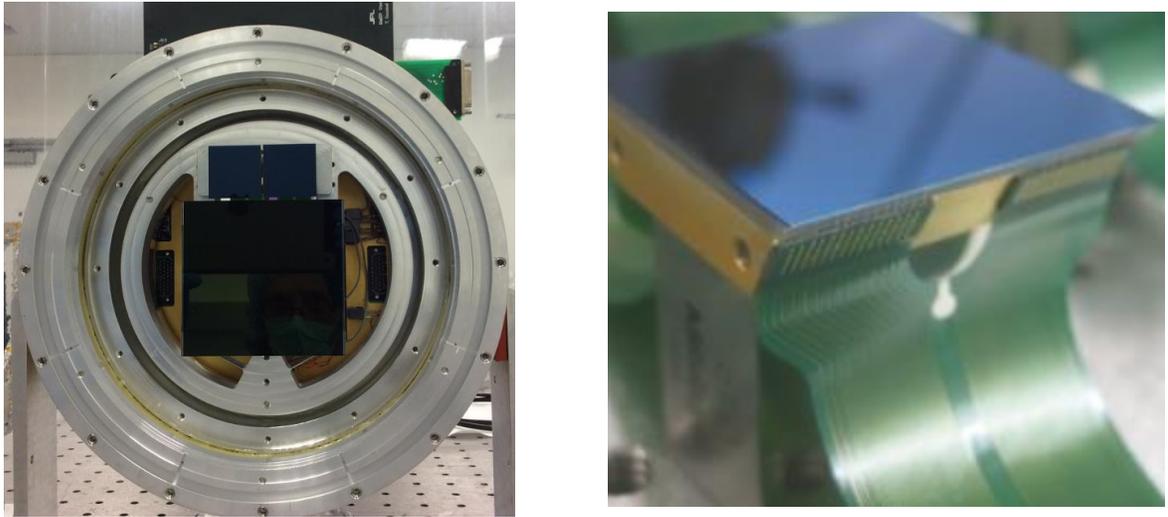

(a) (b)

**Figure 4.9**. (a) The WaSP Guide and Focus CCDs installed in their dewar together with the science detector. The two 2D-doped and AR coated parts are distinguishable by their blue hue. (b) A side view of one of the WaSP detectors showing the custom packaging in which the support and flex cables to stay within the die footprint, allowing for four-side buttable mounting.

As previously mentioned, our collaborators on both WaSP and ZTF were particularly interested in improving performance in the U-band, but not at the loss of signal in the red. The AR coating development for WaSP, previously described,[59] was developed with this objective in mind. Designs with single and multiple layer films were developed based on $Al_2O_3$, $HfO_2$, $SiO_2$ and $TiO_2$. Test coatings were prepared on test samples as well as functional test devices; test samples were evaluated in terms of their optical performance and related properties (e.g., surface and interface roughness) prior to moving to test devices. Transmittance data (100-R %) from several of the WaSP/ZTF coating development samples are presented in Fig. 4.10, along with the predicted performance based on spectroscopic models. Based on sample test data, design **C** was ultimately selected for WaSP/ZTF. Note that in cases where $HfO_2$ was to be deposited directly onto the silicon surface, a 2-nm $Al_2O_3$ diffusion barrier (d-$Al_2O_3$) was used in order to prevent an unfavorable chemical reaction between $HfO_2$ and silicon.[59,60]

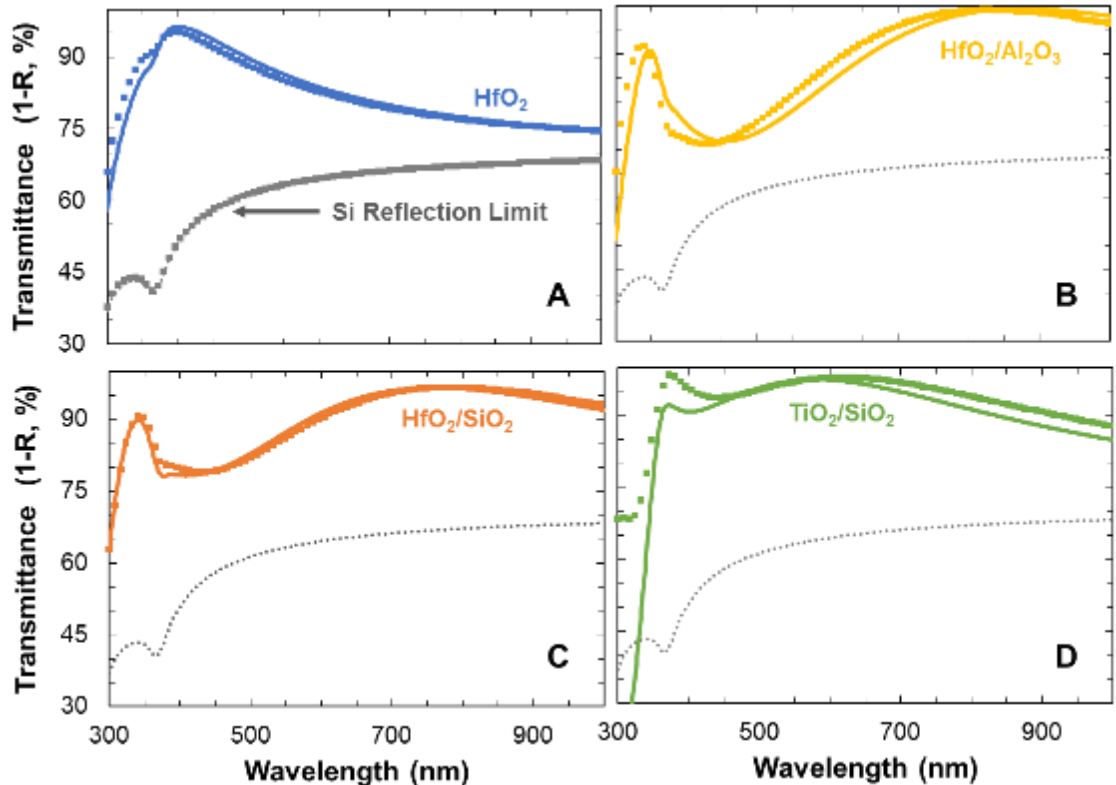

**Figure 4.10.** Model and measured transmittance data (100-R) for WaSP/ZTF test samples (i.e., not test devices). The modeled transmittance data (solid lines) include expected losses due to reflection and absorption, while the measured data (square markers) was calculated based on sample reflectance only. Also shown is the reflection limited response for bare silicon (dotted line) together with a bare silicon reference sample (grey squares, upper left plot). Note samples A, B, and C each included a 2-nm d-$Al_2O_3$.[59]

First light images from the WaSP guide and focus detectors are shown in Fig. 4.11; these composite images (each exposure 100 s) were acquired during commissioning activities with without full waveform optimization.

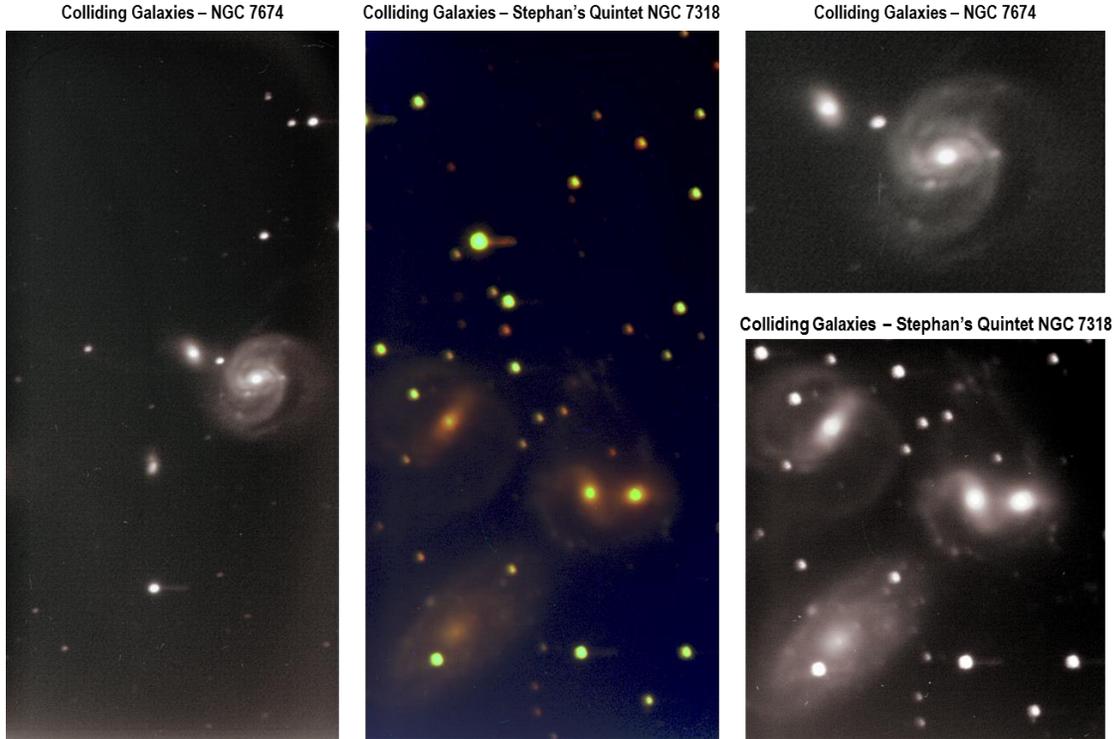

**Figure 4.11**. First light images from WaSP 2D-doped and AR coated guide and focus detector. Shown are composite images comprised on 100-s exposures from the G', R' and U' bands (G'=green, R'=red; U'=blue; color composite created with Photoshop CC 2017).

*4.2.5  Palomar (III): EMCCD*

A 2D-doped 1K×2K EMCCD was processed for on sky observation at Palomar. A single layer AR coating designed for FIREBall-2 (see Sec. 4.1.5) was selected for this device in order to allow higher efficiency beyond the FIREBall narrowband response. Three observation runs at Palomar were completed with this detector behind the Cosmic Web Imager (CWI)[61,62] as part of an ongoing campaign to evaluate the detector's on sky performance and to obtain a science data to set the stage for future space missions. The first of these runs was in Nov. 2015, with follow up in spring and fall of 2016.[57] The detector was operated at -105° C, 10 MHz, initially with the NÜVÜ v.2 CCCP controller and the most recent run with the v.3. CWI is an integral field spectrograph (IFS) with 24 slices that cover a 60 x 40 arcsec$^2$ area with a resolution R~5000. The EMCCD is significantly smaller than the normal CWI detector, and for these runs only half the field of view was observed (12 out of 24 slices, or 30×40 arcsec$^2$). Results from these runs were compared to equivalent observations with the normal CWI detector. We compared both photon counting operation, at a range of gain values, and normal mode operation. The data gathering process has involved many short integrations on sky to track sky lines, and hence flexure during a long exposure; this will allow for higher resolution flexure correction. In addition, these runs were used to verify measured QE results using standard stars, in both normal mode and EM operation. Figure 4.12 shows the results of the second run at Palomar behind the CWI.

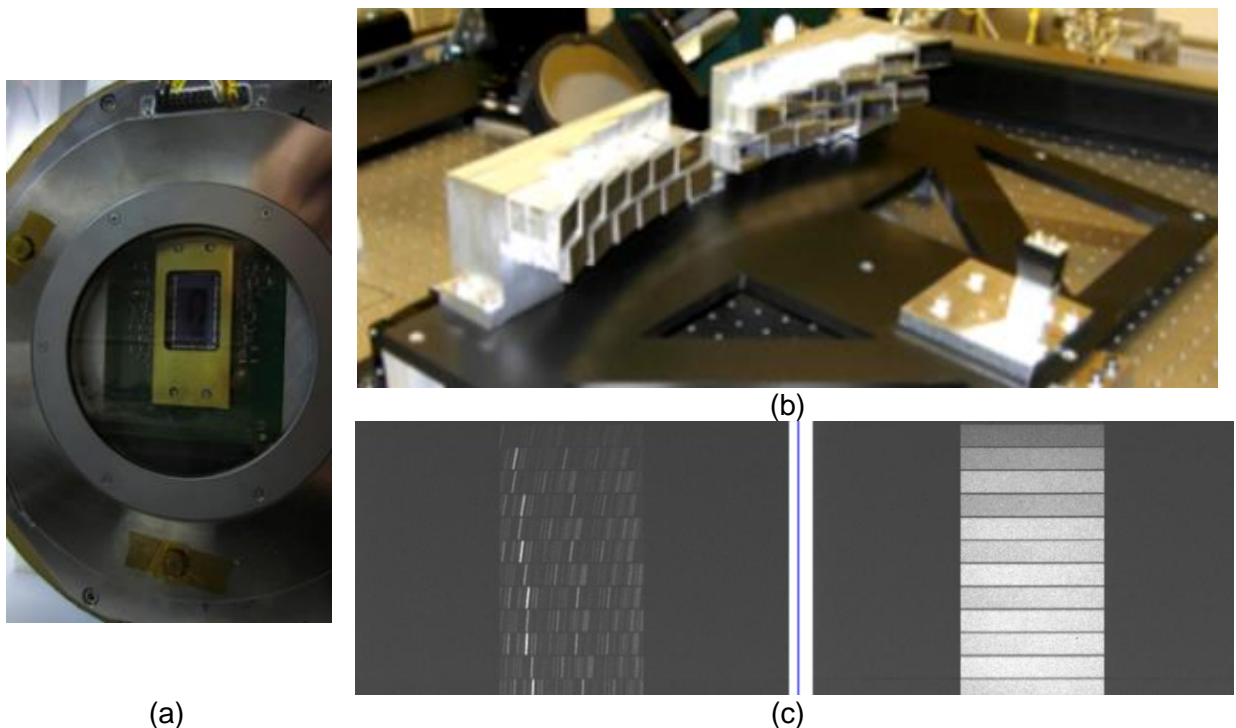

**Figure 4.12**. Photographs of (a) the FIREBall-2 detector prototype installed in the dewar for measurements at (b) CWI. (c) Test images acquired with the 2D-doped, AR-coated EMCCD from CWI. The tiled image shows arcflat and continuum flat images needed for calibration geometry. These 5-s exposures were acquired using the EM/gain function, which improves signal and significantly shortens calibration time.

*4.3 Industry-sponsored Efforts as Relevant to Space-based Applications*

Part of the innovation and development described here is highly leveraged by partnership and work with industry. The high stability, high durability, and high efficiency response of 2D-doped silicon detectors has found applications in semiconductor industry; a few examples are described below.

*4.3.1 Solar-Blind APDs*

One such project is a collaboration between JPL, Radiation Monitoring Devices, Inc. (RMD) Caltech (Prof. D. Hitlin) on the development of gamma ray scintillation detectors with sub-nanosecond temporal resolution and the capability to withstand unprecedented rates and doses of high energy gamma radiation.[63,64] The system consists of doped $BaF_2$ scintillating crystals and solar-blind silicon APDs for detecting the fast scintillation component of $BaF_2$. High efficiency and fast response are achieved by superlattice-doping of the APD. Integrated solar-blind MDFs enable efficient detection of the fast component of $BaF_2$ scintillation at 220 nm, with strong suppression of the slow component at 330 nm (Fig. 4.13). This work represents a significant move toward solar-blind silicon.

The ultrafast response time of these detectors is ideal for NASA applications such as X-ray pulsar navigation, time-gated Raman spectroscopy, and planetary gamma ray spectrometers. NASA has flown a number of instruments and missions using scintillation detectors, including MESSENGER's Gamma ray and neutron spectrometer; the CGRO Energetic Gamma Ray Experiment Telescope (CGRO/EGRET); the Fermi Large Area Telescope (Fermi/LAT); and the Fermi Gamma-Ray Burst Monitor (Fermi/GBM). The unique capabilities of this sub-nanosecond scintillation detector are an enabling technology for NASA's Advanced Compton Telescope and other missions requiring high efficiency gamma-ray detection with excellent time resolution, such as a follow-on to the Fermi-LAT.

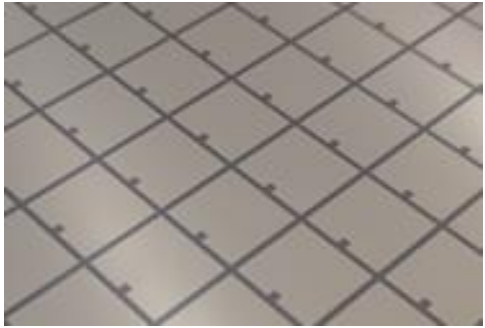
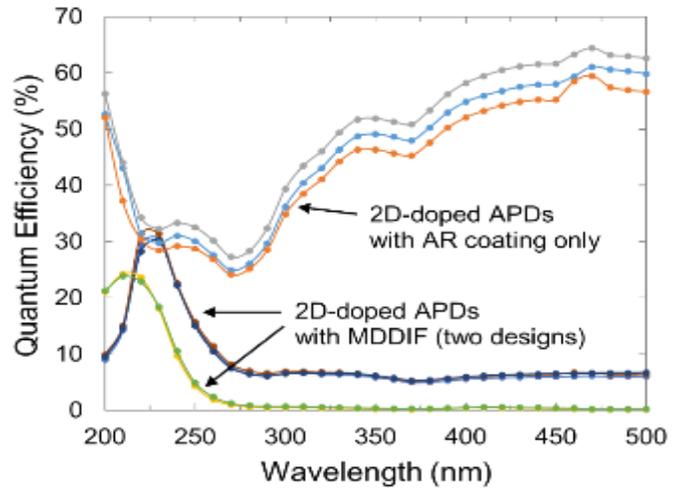

(a)  (b)

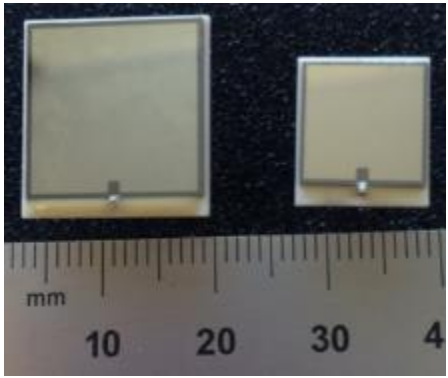
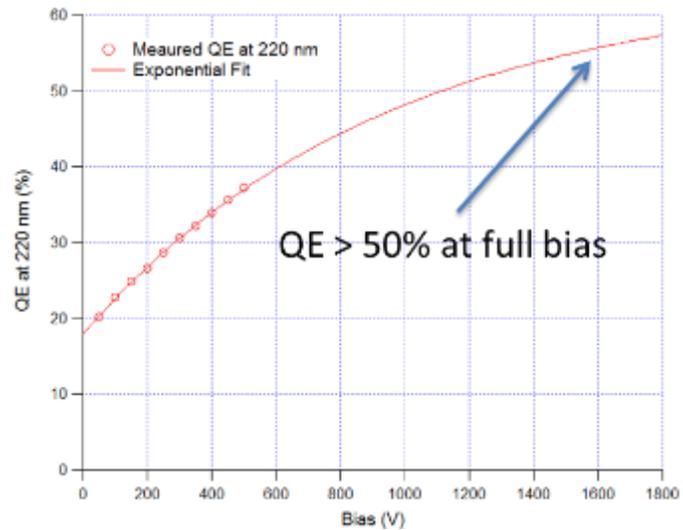

(c)  (d)

**Figure 4.13**. (a) Close-up view of a fully processed APD wafer following superlattice doping, deposition and patterning of the MDF. (b) Response of 2D-doped APDs with alternate integrated MDF designs for the 200-240 nm range show comparable in-band QE, and good out-of-band rejection compared to devices with AR coating only. (c) Packaged APDs of different sizes. (d) The QE at 220 nm plotted as a function of bias voltage for a 2D-doped APD with integrated MDF.

*4.3.2 Ultra-stable CMOS Imagers*

Many scientific and commercial detectors are exposed to ionizing radiation. In such environments, detectors can rapidly lose sensitivity, develop instabilities, and in extreme cases, cease to function altogether. One such application is wafer and reticle inspection in semiconductor foundries. As the industry has progressed toward higher density circuits, state-of-the-art metrology systems have evolved to use pulsed, deep ultraviolet (DUV) lasers for optical detection of defects on the size scale of tens of nanometers. Unfortunately, state-of-the-art detectors degrade rapidly under intense illumination with DUV lasers. To address this challenge, JPL and Alacron have recently developed a high performance, DUV camera with a superlattice-doped CMOS imaging detector. Superlattice-doped detectors are uniquely stable against surface damage caused by DUV photons, as demonstrated by lifetime tests comprising exposure to pulsed, DUV lasers at 263 nm and 193 nm. Superlattice-doped CMOS imaging detectors survived long-term exposure to more than two billion laser pulses, with no measurable degradation, no blooming, and no persistence. Additional details regarding this work are available in Hoenk et al. 2013.[25] The stability demonstrated with the 2D-doped CMOS platform has immediate application in precision photometry for astrophysics and cosmology. Our current focus on 2D-doped, low-noise CMOS will build on the work presented here to benefit future space missions.

*4.4 Satellite Deployment*

The end-to-end post fabrication processing which includes 2D passivation and has been described in this article has been applied to a variety of device designs and formats. These devices have been extensively characterized by various groups and laboratories and have been deployed in suborbital experiments and observatories. Examples of orbital satellite missions that are using these devices in their baseline design are described here.

*4.4.1 Mission of Opportunity: ULTRASAT*

The Ultraviolet Transient Astronomy Satellite (ULTRASAT) is a partnership mission of opportunity concept that will further our understanding of the transient UV sky. ULTRASAT will undertake a wide-field UV time-domain survey of the sky, and provide near UV light curves for hundreds of young supernovae. Both WaSP and ZTF (Sec. 4.2.4) serve as science pathfinders for ULTRASAT, as they will be performing similar measurements, albeit in a different portion of the spectrum. An additional goal of ULTRASAT is to constrain the physics of the sources of gravitational waves through a search for UV emission from a large sample of events. ULTRASAT is unique in its capability to rapidly observe a large fraction of the sky with its large field of view for sky localization of GW sources. This enables ULTRASAT to detect or constrain UV emission from neutron star mergers.

In order to achieve its objectives ULTRASAT requires a large field of view, high sampling cadence and high sensitivity. The ULTRASAT payload consists of a modified Schmidt telescope and a UV camera (ULTRACam) together providing a >210 deg$^2$ field of view and 21.3 AB limiting magnitude in the ULTRASAT band (220-280 nm). The ULTRACam focal plane is a 40-megapixel mosaic of five 2D-doped and AR coated 2K×4K CCDs (e2v CCD 42-80) optimized for NUV (Fig. 4.14). Work on the WaSP/ZTF and FIREBall-2 projects serve as technology pathfinders for the ULTRACam detector.

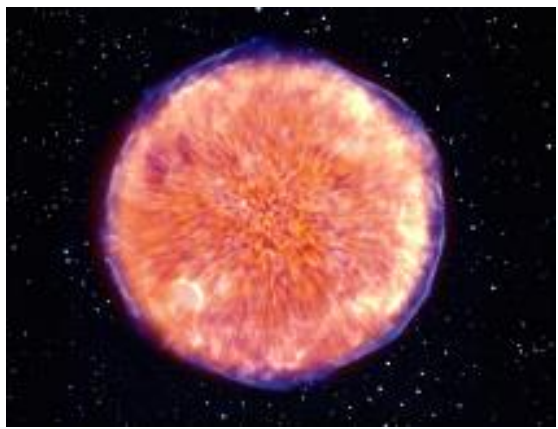 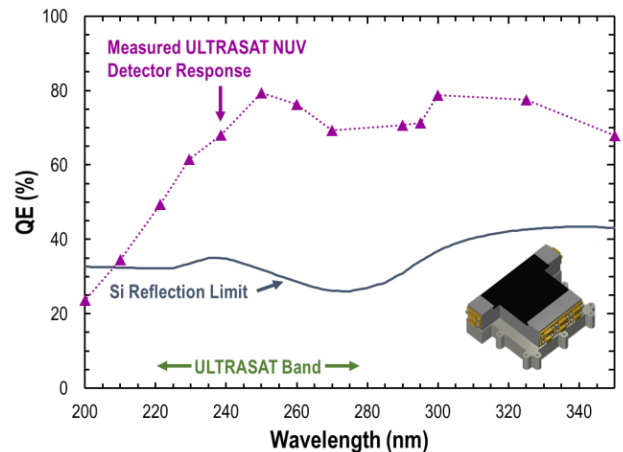

(a) (b)

**Figure 4.14**. (a) One of ULTRASAT's objectives is to study the death of massive stars. Shown in violet is the shock breakout of the star, which is expected to shine strongly in the UV. (b) Plots showing the measured QE for the ULTRASAT NUV detector (purple triangles, dotted line), a 2D-doped and AR coated detector. Also shown is the silicon reflection limit, that is, the response of a non-AR coated, 2D-doped CCD. Inset: concept of the ULTRASAT focal plane with a mosaic of five 2k×4k CCDs.

*4.4.2 Explorer Class Mission: Mapping the Circumgalactic and Intergalactic Medium*

An explorer-class mission concept is currently under development that will map Lyman-α and metal emission lines from the circumgalactic medium (CGM) of low redshift galaxies, similar to the narrower search conducted by FIREBall-2 but over a larger wavelength range. The satellite will use a wide-field UV imaging spectrometer, and the mission has baselined an array of delta-doped EMCCDs for a two channel UV IFS (near UV and far UV). These observations will link the evolution of galaxies and star formation through cosmic time with the evolution of gas flows into and out of the CGM and IGM. The primary goal is to explain the dramatic evolution of cosmic star formation in the modern age and the formation of the galaxies we see today. The mission will survey the CGM in emission, characterizing a reservoir of the majority of the baryons in the universe. It will search for emission from the general IGM in deep observations. UV capability is required for these observations because the UV sky is very dark and most emission is in rest-frame UV lines that appear in the space UV for redshift 0-1.5. The focal plane consists of an array of detectors with customized AR-coatings dictated by their spectral position in the array, covering 120-350 nm between the two channels. In the past several years we have demonstrated >50% QE over individual bands throughout the 1200-3000 Å range (Fig. 4.15). Our work on the FIREBall-2 mission serves as a pathfinder for both the CGM/IGM science as well as the detector technology (i.e., the 2D-doped and AR coated EMCCD).[55,56,65]

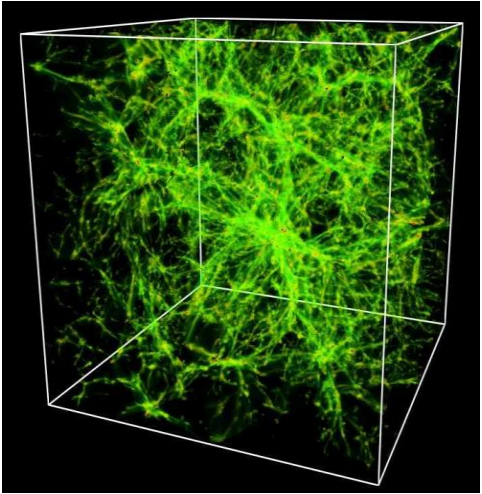 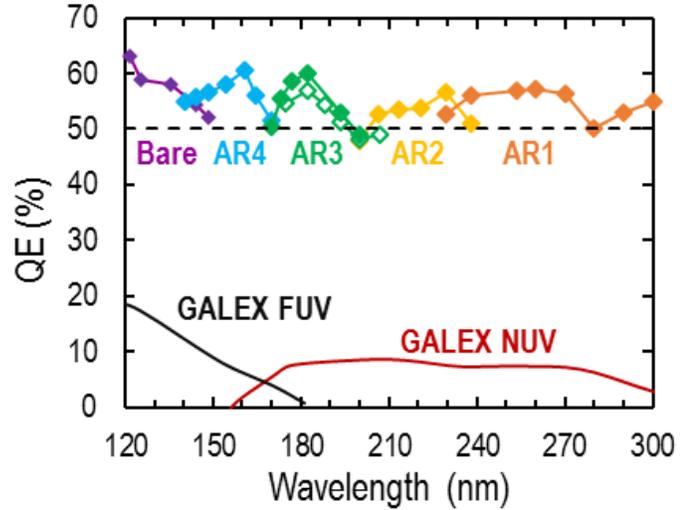

(a) (b)
**Figure 4.15.** (a) Studies of CGM/IGM will shed light on the evolution of galaxies and star formation through cosmic time. (b) Early efforts for this mission have demonstrated >50% QE throughout the spectral range of interest using AR-coated, 2D-doped detectors. For this work each spectral region was targeted using a customized AR coating. Both conventional CCDs (closed diamonds) and EMCCDs (open diamonds) were used in this work.[66] Responses of the GALEX MCP detectors are shown for comparison.[67,68]

*4.5 Looking to the Future*

*4.5.1 Probe Class Mission: ANUBIS*

The Astronomical UV Probe Imager Spectrograph (ANUBIS) is a probe-class UV-Optical Space Observatory currently in the formulation phase. The initial mission concept for ANUBIS combines a wide field UV-optical imager with a FUV spectrograph. This next generation observatory will be capable of conducting wide-field imaging surveys to study the formation and survival of stellar and planet forming environments. ANUBIS will include a FUV spectrograph with the wavelength coverage to reach the forest of diagnostic emission and absorption lines necessary to study the interface between galaxies and the intergalactic medium, to probe the structure and dynamics of the interstellar medium in all its phases, both locally and in extragalactic systems.

Imaging surveys of large angular areas are necessary to conduct assays of widely scattered populations of objects—requiring large angular area coverage to locate the sources and high angular resolution to characterize sources once found. To maximize survey efficiency, multiple imaging channels are needed, fed by dichroics, and each focal plane must be optimized for its particular passband. This requires the production of large numbers of detectors, as well as the capability to tune the response of the detector to specific passbands.

ANUBIS employs a dichroic design with two large-area FPAs, splitting the spectral range into 200-510 nm and 500-1000 nm "blue" and "red" channels. To achieve the highest QE in this range, ANUBIS baselines JPL's 2D doping in combination with high-purity CCD arrays. Two different AR coatings are designed to cover the blue and red channels of ANUBIS (Fig. 4.16). A collaborative JPL-ASU effort has demonstrated the QE for these two channels. In addition to the high QE, photometric stability that is provided by the 2D doping is essential for achieving ANUBIS objectives.

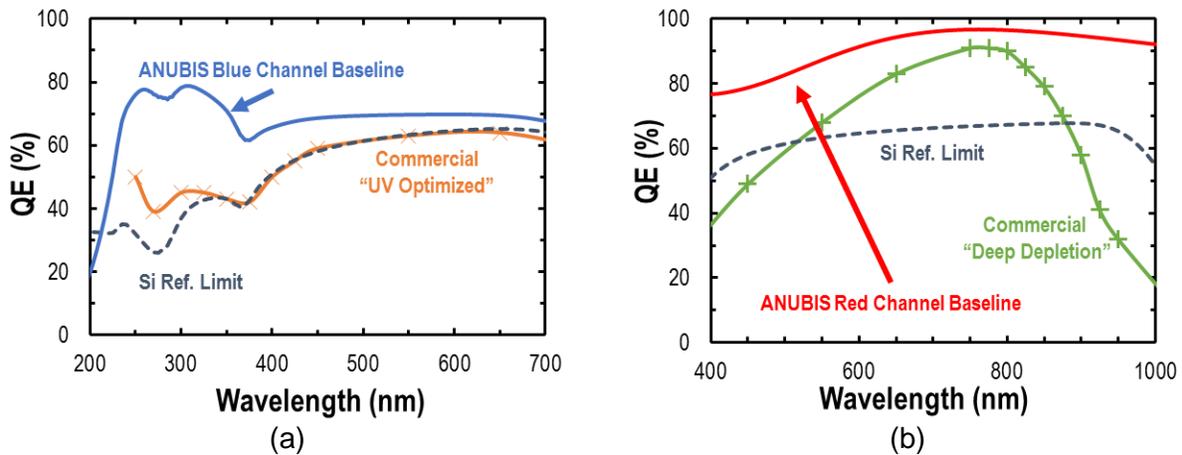

**Figure 4.16**. The QE models for ANUBIS (a) blue and (b) red channels. For comparison, the QE of commercially available UV-optimized (orange) and Deep Deletion (green) CCDs are also plotted, as is the silicon reflection limit (grey).

Additionally, 2D-doped EMCCDs can be optimized for the ANUBIS FUV spectrograph, offering photon counting, high UV efficiency, and additional out of band rejection.

*4.5.2  LUVOIR and HabEx*

NASA is currently studying four Flagship-class mission concepts as input to the 2020 National Research Council Decadal Survey on Astronomy and Astrophysics. Two of those mission concepts involve the use of the UV-optical passbands to address both general astrophysics and exoplanet direct imaging. LUVOIR is currently being conceived as either a 9-m or 16-m aperture segmented telescope, while HabEx is being considered as either a 4-m monolith or a 6.5-m segmented telescope. In both cases, wide field imagers are being baselined as foundational instruments together with FUV spectrographs to address the general astrophysics science goals of each mission concept.

In the case of the wide field imager designs being formulated, the instrument calls for a large number of pixels to adequately sample the point spread function (PSF) of the telescope, while providing enough discovery space (AΩ) to perform efficient surveys in combination with characterization and analysis. The mirror coatings being considered will provide high reflectivity down to 200 nm; thus, the detectors chosen will need to deliver an optimum performance across the 200-1000 nm passband, with some possibility of dichroic splits to maximize efficiency. To this end, detector technologies will be needed that can be treated to deliver the needed response with a minimal level of risk. The 2D-doped, high resistivity detectors delivered for WaSP/ZTF and CHESS, as well as those baselined for ANUBIS, provide the required spectral sensitivity in a proven platform.

The FUV spectrographs will similarly likely require large focal planes due to the need to provide a multi-object spectroscopic (MOS) capability; this will likely require FPA's measured in the 10's of centimeters. Detectors sensitive from the NUV (300 nm) down as low as 90 nm, depending on the mirror coating choices made, will be required. The selected focal plane technologies will need to deliver an efficient response across that passband at a resolution element pitch of probably 10-20 μm in size. The FUV-optimized, 2D-doped detectors that have been delivered for various suborbital missions, including FIREBall-2, as well as those under development for ULTRASAT, the CGM/IGM Mapper, and ANUBIS, are ideal candidates.

The promise of these next generation mission concepts is staggering when compared with modern facilities even in the 30-m class being considered for the ground, so the detectors chosen to baseline the performance will need to represent the very best and most cutting edge technologies the community can provide. As such to be eligible for adoption the requisite technologies will need to demonstrate TRL 6 or 7 performance by the end of Phase A which we expect to be no earlier than 2024, assuming WFIRST-AFTA is launched as projected. The development and deployment of high performance 2D-doped silicon arrays described here put us in a strong position for the future.

## 5 Conclusions

We have presented an overview of the development of high efficiency silicon detectors, from the technology invention and laboratory demonstration of the first 100×100-pixel array measured to exhibit 100% internal QE in the NUV to Visible to today's 12-megapixel arrays and multiple arrays produced using our high throughput processing. We have shown a survey of deployment of these 2D detectors in space through suborbital experiments and in ground based observatories. Along the way, major milestones in UV/Optical/NIR band have been demonstrated including but not limited to high QE in n-channel as well as p-channel silicon arrays; 2D-doped, high QE electron multiplying CCDs; 2D doping of CMOS arrays; development of detector-integrated filters with out-of-band rejection ratios rivaling those of microchannel plates but with higher efficiency. Adding to that, the scaled-up manufacturing of these detectors allows population of large FPAs foreseen for the next generation of space telescopes such as LUVOIR and HabEx.

**Acknowledgements**

The research was carried out in part at the Jet Propulsion Laboratory, California Institute of Technology, under a contract with the National Aeronautics and Space Administration. The authors gratefully acknowledge the support from NASA-SAT, NASA-APRA, JPL' RTD program, the Keck Institute for Space Studies, and many years of other support from NASA and JPL and other agencies. E.T.K acknowledges the support by Nancy Grace Roman Fellowship, the Millikan Prize in Experimental Physics, and the NSF Fellowship. The authors would like to acknowledge generous and excellent collaborative support from e2v including A. Reinheimer, P. Jerram, P. Pool, P. Jorden, and P. Fochi on the 2-D doped EMCCDs. The authors gratefully acknowledge the collaborative work with LBNL's S. Holland, C. Bebek, N. Roe and their team that has led to the demonstration and development of 2D-doped, p-channel CCDs; as well as Bruno Milliard and Robert Grange of LAM for the FIREBall-2 spectrograph data. We acknowledge M. McClish of RMD for detector QE data and wafer processing. S.R. Kulkarni, R. Smith, and J. Milburn of Caltech Optical Observatories and Palomar Observatory for WaSP delta-doped array first-light images.

...